%% file: main.tex
\documentclass[twoside]{article}

\usepackage[accepted]{aistats2022}

\usepackage{enumitem}
\input{packages}

\newboolean{showcomments}
\setboolean{showcomments}{true}
\newcommand{\sj}[1]{  \ifthenelse{\boolean{showcomments}}
{ \textcolor{red}{(SJ says:  #1)}} {}  }
\newcommand{\vinay}[1]{  \ifthenelse{\boolean{showcomments}}
{ \textcolor{red}{(Vinay says:  #1)}} {}  }
\newcommand{\jv}[1]{  \ifthenelse{\boolean{showcomments}}
{ \textcolor{red}{(JV says:  #1)}} {}  }
\newcommand{\rs}[1]{  \ifthenelse{\boolean{showcomments}}
{ \textcolor{red}{(RS says:  #1)}} {}  }
\newcommand{\dm}[1]{  \ifthenelse{\boolean{showcomments}}
{ \textcolor{red}{(DM says:  #1)}} {}  }
\newcommand{\sd}[1]{  \ifthenelse{\boolean{showcomments}}
{ \textcolor{red}{(SD says:  #1)}} {}  }

\newtheorem{theorem}{Theorem}

\newtheorem{lemma}[theorem]{Lemma}

\DeclareMathOperator*{\argmax}{argmax}

\DeclareMathOperator*{\eqdef}{\stackrel{\text{\tiny def}}{=}}

\usepackage{dblfloatfix}

\usepackage[round]{natbib}

\begin{document}

%

%
\runningauthor{Jain, Shah, Gupta, Mehta, Nair, Vora, Khyalia, Das, Ribeiro, Kalyanakrishnan}

\twocolumn[

\aistatstitle{PAC Mode Estimation using PPR Martingale Confidence Sequences}

\aistatsauthor{ Shubham Anand Jain$^*$\footnotemark[1] \And Rohan Shah$\dagger$\footnotemark[1] \And  Sanit Gupta$\dagger$\footnotemark[2] \And Denil Mehta$\dagger$\footnotemark[2]}
\aistatsauthor{
Inderjeet Nair$\dagger$\footnotemark[2] \And  Jian Vora$^*$\footnotemark[2] \And Sushil Khyalia$\dagger$ \And Sourav Das$^\ddagger$}


\aistatsauthor{Vinay J. Ribeiro$\dagger$ \And Shivaram Kalyanakrishnan$^\dagger$}

\aistatsaddress{$^\dagger$Indian Institute of Technology Bombay, $^*$Stanford University, $^\ddagger$ University of Illinois Urbana-Champaign\\
(Authors marked ${}^1$ contributed equally; authors marked ${}^2$ contributed equally.)}



]

\input{abstract}

\input{sections/introduction}

\input{sections/pprmcs}

\input{sections/ppr-me}

\input{sections/theory}


\input{sections/applications}

\input{sections/conclusion}


\input{sections/acknowledgements}

\bibliography{references}

\clearpage
\appendix 

\thispagestyle{empty}

\onecolumn \makesupplementtitle

\input{appendices/implementationdetails}

\input{appendices/code}

\input{appendices/1v1and1vrtermination}
\input{appendices/optimality_final}

\input{appendices/proofoflemma3}

\input{appendices/dcb}

\input{appendices/bihar-table}






\end{document}

%% file: packages.tex
\usepackage{ifthen}
\usepackage{amsmath}
\usepackage{amssymb}       
\usepackage{amsthm} 

\usepackage{verbatim}
\usepackage{booktabs}
\usepackage{multirow}
\usepackage{siunitx}

\usepackage{hyperref}

\usepackage{titlesec}

\usepackage{graphicx}
\usepackage{float}
\usepackage{multicol}
\usepackage{dsfont}

\usepackage{caption, subcaption}

\usepackage{url}

%% file: abstract.tex
\begin{abstract}
We consider the problem of correctly identifying the \textit{mode} of a discrete distribution $\mathcal{P}$ with sufficiently high probability by observing a sequence of i.i.d. samples drawn from $\mathcal{P}$. This problem reduces to the estimation of a single parameter when $\mathcal{P}$ has a support set of size $K = 2$. After noting that this special case is tackled very well by prior-posterior-ratio (PPR) martingale confidence sequences \citep{waudby-ramdas-ppr}, we propose a generalisation to mode estimation, in which $\mathcal{P}$ may take $K \geq 2$ values. To begin, we show that the ``one-versus-one'' principle to generalise from $K = 2$ to $K \geq 2$ classes is more efficient than the ``one-versus-rest'' alternative. We then prove that our resulting stopping rule, denoted PPR-1v1, is asymptotically optimal (as the mistake probability is taken to $0$). PPR-1v1 is parameter-free and computationally light, and incurs significantly fewer samples than competitors even in the non-asymptotic regime.
We demonstrate its gains in two practical applications of sampling: election forecasting and verification of smart contracts in blockchains. 
\end{abstract}

%% file: sections/introduction.tex
\section{INTRODUCTION}
\label{sec:intro}

We investigate the problem of estimating the mode of a given, arbitrary, discrete probability distribution $\mathcal{P} = (p, v, K)$ by observing a sequence of i.i.d. samples drawn according to $\mathcal{P}$. Here $\mathcal{P}$ takes values from the support set $v = \{v_{1}, v_{2}, \dots, v_{K}\}$ according to the probability vector $p = \{p_1, p_2, \dots, p_K\}$ for some $K \geq 2$. For $1 \leq i \leq K$, the probability of obtaining $v_{i}$ from $\mathcal{P}$ is $p_{i}$. We assume that $\mathcal{P}$ has a unique mode, and without loss of generality, $p_{1} > p_{2} \geq p_{3} \geq p_{4} \geq \dots \geq p_{K}$ (which makes $v_{1}$ the mode).

Our aim is to provide a procedure $\mathcal{L}$ to identify the mode of $\mathcal{P}$. At each step $t \geq 1$,
$\mathcal{L}$ can either ask for a sample $x^{t} \sim \mathcal{P}$ or it can terminate and declare its answer. For ``mistake probability'' $\delta \in (0, 1)$, $\mathcal{L}$ is said to be $\delta$-correct if for every qualifying discrete distribution $\mathcal{P}$, $\mathcal{L}$ terminates with probability $1$ and correctly identifies the mode of $\mathcal{P}$ with probability at least $1 - \delta$. If $\mathcal{L}$ terminates after observing the sequence of samples $x^{1}, x^{2}, \dots, x^{T}$ for some $T \geq 1$, we may assume that its answer is the most frequent value of $\mathcal{P}$ in this sequence, since it can be argued that no other choice can decrease the mistake probability across all problem instances. Hence, it is convenient to view $\mathcal{L}$ simply as a \textit{stopping rule}, which only needs to decide when to terminate. We aim to devise a $\delta$-correct stopping rule $\mathcal{L}$ with low \textit{sample complexity}---informally the number of samples $T$ observed before stopping. 

In order to make 
our problem ``properly'' PAC, we could introduce a tolerance parameter $\epsilon$, with the implication that any returned value with associated probability at least $p_{1} - \epsilon$ will be treated as correct. We omit this generalisation, noting that 
it can be handled quite easily by the methods proposed in the paper. In fact, our version with $\epsilon = 0$ exactly matches the problem defined by \citet{Shah_Choudhury_Karamchandani_Gopalan_2020}, whose state-of-the-art results are our primary baseline. \citet{Shah_Choudhury_Karamchandani_Gopalan_2020} show the following lower bound.

\begin{theorem}[Lower bound~\citep{Shah_Choudhury_Karamchandani_Gopalan_2020}]
\label{thm:lowerbound}
Fix $\delta \in (0, 1)$, $K \geq 2$, and a $\delta$-correct stopping rule $\mathcal{L}$. For each categorical distribution $\mathcal{P} = (p, v, K)$, the expected number of samples observed by $\mathcal{L}$ is at least $$\text{LB}(\mathcal{P}, \delta) \eqdef \sup_{\mathcal{P}^{\prime}:\text{mode}(\mathcal{P}^{\prime}) \neq \text{mode}(\mathcal{P})} \frac{1}{\text{KL}(\mathcal{P}||\mathcal{P}^{\prime})}\ln\left(\frac{1}{2.4\delta}\right),$$
where $KL(P||P')$ denotes the KL divergence between categorical distributions $P$, $P^{\prime}$ with same support set.
\end{theorem}
\citet{Shah_Choudhury_Karamchandani_Gopalan_2020}
also give a stopping rule, denoted $\mathcal{A}_{1}$, whose sample complexity 
is upper-bounded to within a logarithmic factor of this lower bound. 

In this paper, we approach the PAC mode estimation problem from a different perspective. In recent work, \citet{waudby-ramdas-ppr} propose prior-posterior-ratio (PPR) martingale confidence sequences as a novel framework to obtain ``anytime'' confidence bounds on unknown parameters of a 
probability distribution. The resulting stopping rule is simple, with no need for tuning, and yet works surprisingly well in practice. Encouraged by this empirical finding, we investigate the application of the PPR martingale test to PAC mode estimation. Below we summarise the contents and contributions of our paper.

\begin{itemize}[leftmargin=*]
\setlength{\itemsep}{0em}
\item We begin by reviewing the 
PPR martingale test~\citep{waudby-ramdas-ppr} in Section~\ref{sec:pprmcs}, and apply it directly to our ``base case'' of $K = 2$. Empirical comparisons establish clear evidence of the relative efficiency of this test.

\item In Section~\ref{sec:ppr-me}, we propose three natural methods to generalise the PPR martingale test to mode estimation ($K \geq 2$). Two of these are the ``one-versus-one'' (1v1) and ``one-versus-rest'' (1vr) approaches used commonly in multi-class machine learning tasks; the third applies the multi-dimensional (MD) variant of the PPR martingale test~\citep{waudby-ramdas-ppr}. The ``one-versus-one'' method, denoted PPR-1v1, is parameter-free, easy to implement, and computationally lighter than competitors. Experiments indicate that PPR-1v1 is also the most sample-efficient among the algorithms.

\item In Section~\ref{sec:theory}, we provide two theoretical arguments to explain the efficiency of PPR-1v1.

\begin{enumerate}[leftmargin=*]
\item We prove that for many commonly used Chernoff bounds, the 1vr adaptation to mode estimation cannot terminate before the 1v1 variant; additionally the MD variant of PPR cannot terminate before PPR-1v1. These results hold for \textit{every single run}, and establish 1v1 as a clear choice for mode estimation. Even $\mathcal{A}_{1}$, originally implemented as a 1vr variant~\citep{Shah_Choudhury_Karamchandani_Gopalan_2020}, is seen to perform much better by switching to 1v1 (although it remains inferior to PPR-1v1).

\item We prove that PPR-1v1 is \textit{asymptotically optimal}, in the sense that for every categorical distribution $\mathcal{P}$, the ratio of the expected sample complexity of PPR-1v1 and $\text{LB}(\mathcal{P}, \delta)$ goes to $1$ as the mistake probability $\delta$ is taken to $0$. To the best of our knowledge, this guarantee is the first of its kind for mode estimation, although similar results have been provided in the multi-armed bandits literature~\citep{pmlr-v49-garivier16a}. Interestingly, 1vr variants (such as $\mathcal{A}_{1}$) appear not to be asymptotically optimal.
\end{enumerate}

\item Over the years, the mode estimation problem has received attention in many different contexts~\citep{Parzen62,Motwani02}. In Section~\ref{sec:practicalapplications}, we illustrate the relevance of PPR-1v1 in two contrasting real-life applications. First, we show that when used as a subroutine, PPR-1v1 can reduce the sample complexity of winner-forecasting in indirect elections~\citep{karandikar2018}. Thereafter, we present its application to
probabilistic verification in permissionless blockchains~\citep{sourav2019}.
\end{itemize}

In short, our paper proposes PPR-1v1 as a novel stopping rule for PAC mode estimation, and provides both theoretical and empirical reasons to justify the choice.

%% file: sections/pprmcs.tex
\section{THE PPR MARTINGALE TEST}
\label{sec:pprmcs}

In this section, we consider the ``base case'' of mode estimation, in which $\mathcal{P}$ takes exactly $K = 2$ values.  Notice that $\mathcal{P}(p, v, 2)$ is a Bernoulli distribution that generates $v_{1}$ with probability $p_{1}$ and $v_{2}$ with probability $p_{2} = 1 - p_{1}$. Treating $p_{1} \in [0, 1]$ as the sole parameter of the distribution, our task is to devise a $\delta$-correct stopping rule to test if $p_{1} > \frac{1}{2}$. Since $p_{1}$ may be arbitrarily close to $\frac{1}{2}$, it is not possible to decide beforehand how many samples suffice for the test to succeed. An unfortunate consequence of having a \textit{random} stopping time is that it cannot be used directly within concentration inequalities such as Chernoff bounds. Rather, stopping rules invariably go through a union bound over all possible stopping times, dividing the mistake probability $\delta$ among them~\citep{ICML12-shivaram,pmlr-v30-Kaufmann13}. Although there has been progress towards optimising this apportioning of $\delta$~\citep{Jamieson14,garivier2013}, resulting methods still have tunable parameters in their ``decay rates'', which govern the stopping time.

The recent development of ``time-uniform'' or ``anytime'' Chernoff bounds~\citep{Howard20} relieve the experimenter of tedious parameter-tuning. Arising from this line of research is the framework of prior-posterior-ratio (PPR) martingale confidence sequences~\citep{waudby-ramdas-ppr}, which yields a simple, intuitive stopping rule. Although the rule may be applied more widely, we restrict our upcoming 
discussion to the Bernoulli case at hand: that is, to test whether $p_{1} > \frac{1}{2}$.

To apply the PPR martingale framework, we maintain a belief distribution $\pi$ for $p_{1}$ over its range $[0, 1]$, and update $\pi$ according to Bayes' rule as samples are observed. Our aim is still to provide a frequentist guarantee that holds for all possible values of $p_{1}$ ($\delta$-correctness). To this end we must ensure that
the prior distribution $\pi^{0}$ gives non-zero density to all possible values of $p_{1}$. We do so by adopting the uniform prior $\pi^{0}(q) = 1$ for $q \in [0, 1]$. For $t \geq 1$, we update our belief distribution after observing sample $x^{t}$: 
\begin{equation*}
    \pi^{t}(q) = \frac{\pi^{t - 1}(q) \cdot (q)^{\mathbf{1}[x^{t} = v_{1}]} \cdot 
(1 - q)^{\mathbf{1}[x^{t} = v_{2}]}
}{\int_{\rho = 0}^{1} \pi^{t - 1}(\rho) \cdot (\rho)^{\mathbf{1}[x^{t} = v_{1}]} \cdot (1 - \rho)^{\mathbf{1}[x^{t} = v_{2}]} d\rho}.
\end{equation*}
The prior-posterior-ratio (PPR) at $q \in [0, 1]$ is given by $R^{t}(q) = \frac{\pi^{0}(q)}{\pi^{t}(q)}$. \citet{waudby-ramdas-ppr} show that the sequence of sets $(C^{t})_{t = 0}^{\infty}$, where $C^{t} \eqdef \{q: R^{t}(q) < \frac{1}{\delta}\}$, is a $(1 - \delta)$-confidence sequence for $p_{1}$~\citep{waudby-ramdas-ppr}. In other words, we have the ``anytime'' guarantee that 
\begin{equation}
\label{eqn:ppr-correctness}
\mathbb{P}\{\: \exists \: t \geq 0: p_{1} \notin C^{t} \} \leq \delta.
\end{equation}

The correctness of \eqref{eqn:ppr-correctness} is shown by establishing that the PPR evaluated at the \textit{true} parameter value, $p_{1}$, is a martingale, and then applying Ville’s inequality for nonnegative supermartingales~\citep[see Appendix B.1]{waudby-ramdas-ppr}. For our special case of estimating the parameter of a Bernoulli distribution, the belief distribution $\pi^{t}$ and hence the PPR $R^{t}$ assume a convenient form if initialised with the uniform prior.
Suppose the sequence of samples up to time $t$ is $x^{1}, x^{2}, \dots, x^{t}$, which contains $s^{t}_{1}$ occurrences of $v_{1}$ and $s^{t}_{2} = t - s_{1}$ occurrences of $v_{2}$. Then for $t \geq 0$ and $q \in [0, 1]$, we obtain $\pi^{t}(q) = \text{Beta}(q; s^{t}_{1} + 1, s^{t}_{2} + 1)$ (the pdf of a Beta distribution with parameters $s^{t}_{1} + 1$ and $s^{t}_{2} + 1$, evaluated at $q$). We can terminate as soon as the $(1 - \delta)$-confidence sequence on $p_{1}$ does not contain $\frac{1}{2}$. For easy readability, let us define indices $\text{first}(t)$ and $\text{second}(t)$, where $(\text{first}(t), \text{second}(t)) \in \{(1, 2), (2, 1)\}$ satisfies $s^{t}_{\text{first}(t)} \geq s^{t}_{\text{second}(t)}$. We obtain the following simple stopping rule, applied at each time step $t \geq 1$.
\begin{center}
\fbox{
\parbox{0.93\columnwidth}{
\textbf{PPR-Bernoulli}: 
Stop, declare $v_{\text{first}(t)}$ as mode iff
$\phantom{aaaaa}\text{Beta}\left(\frac{1}{2}; s^{t}_{\text{first}(t)} + 1, s^{t}_{\text{second}(t)} + 1\right) \leq \delta.$
}
}
\end{center}

Note that the LHS of the PPR-Bernoulli stopping rule can be evaluated \textit{exactly} as a rational, using integer arithmetic, requiring only a lightweight incremental update after each sample. As we see shortly, many other stopping rules require much heavier computation, such as to perform numerical optimisation.



\subsection{Empirical Comparisons,  \texorpdfstring{$K = 2$}{}}
\label{subsec:empiricalcomparisonk2}

\begin{figure}[b!]
\centering
\subfloat[\centering $p_1$ varied; $\delta = 0.01$. ]{{\includegraphics[width=\columnwidth]{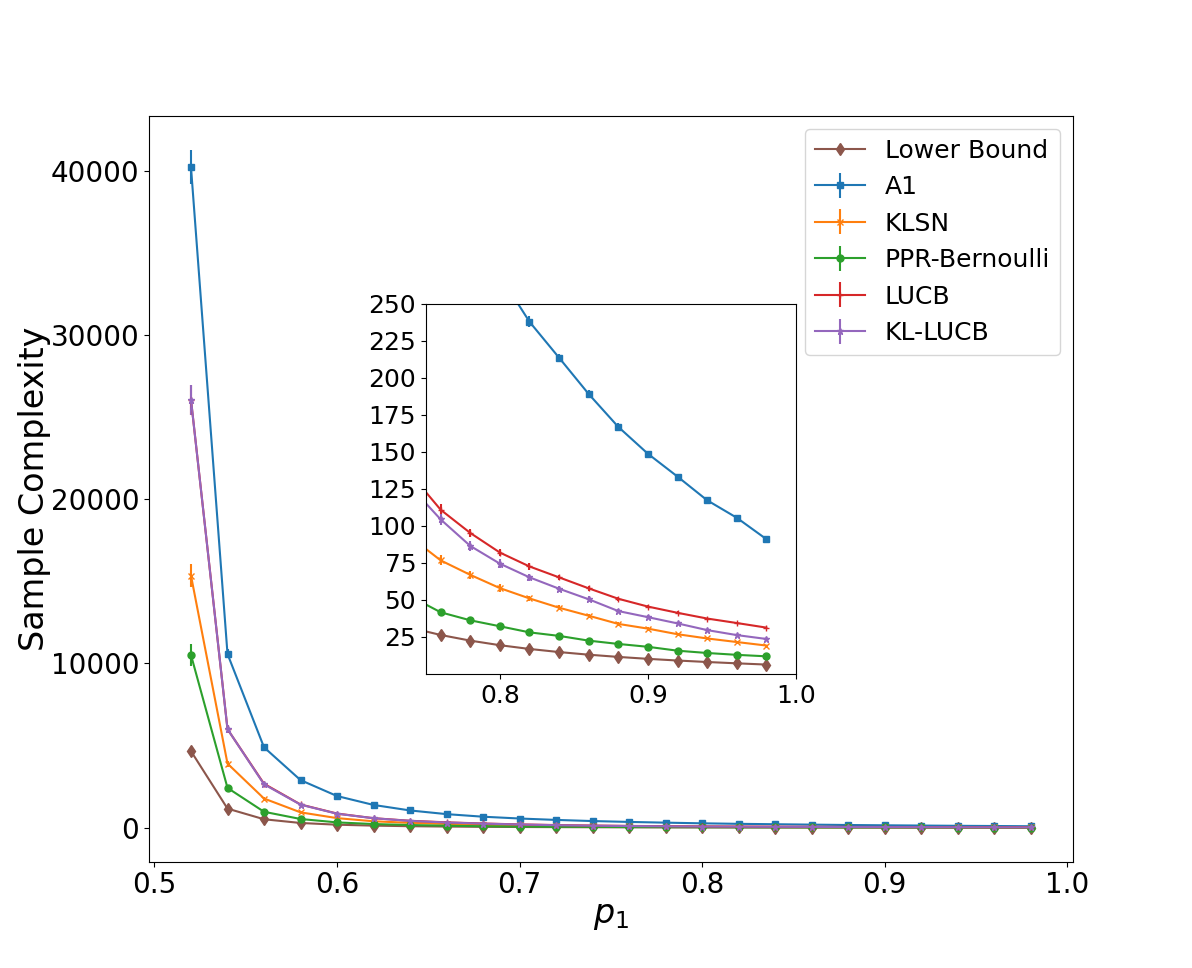} }
\label{fig:bernoulli-pvaried}}

\subfloat[\centering $\delta$ varied; $p_{1} = 0.65$. ]{{\includegraphics[width=\columnwidth]{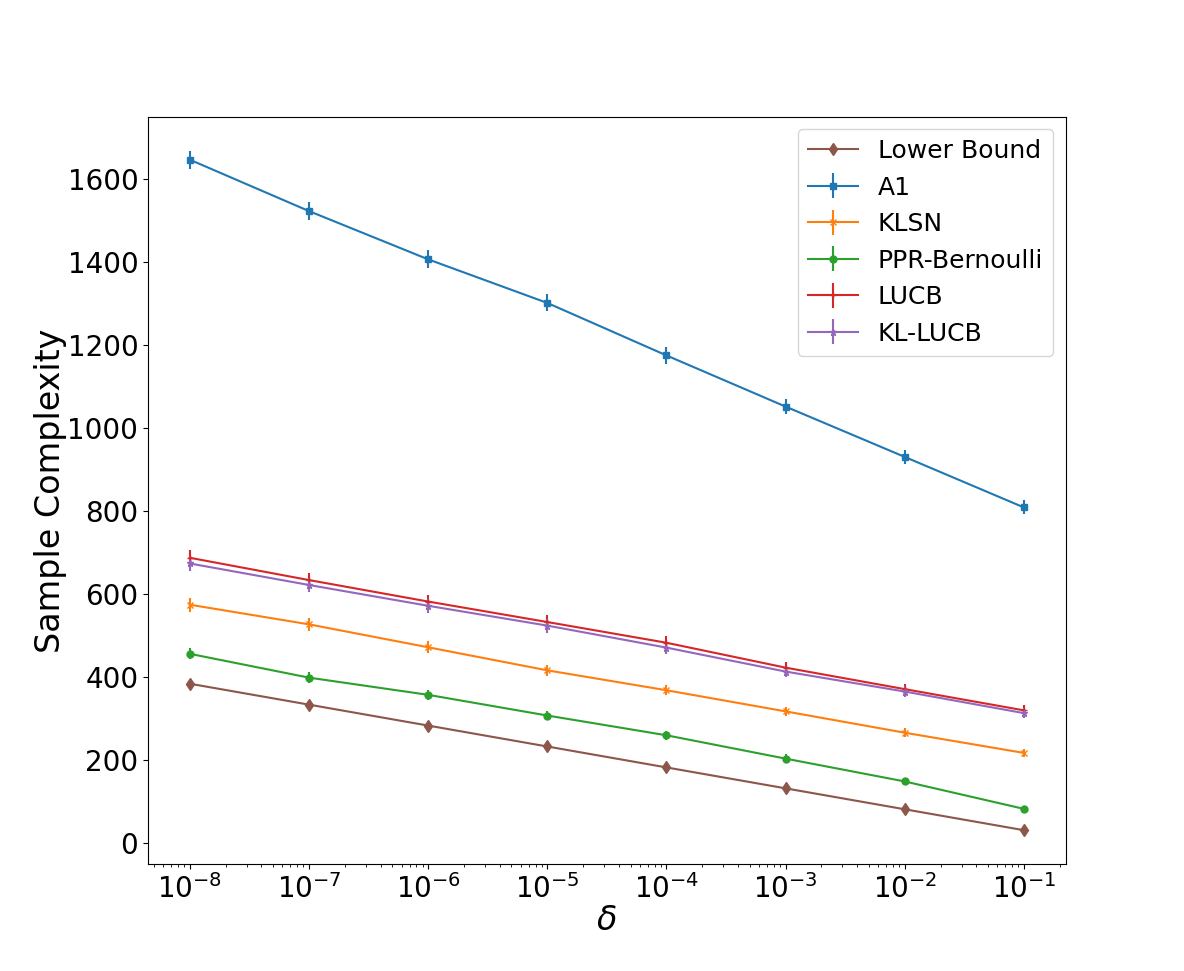} }
\label{fig:bernoulli-deltavaried}
}%
\caption{Comparison of stopping rules for the Bernoulli case ($K = 2$). Both plots show sample complexity: in (a) as $p_{1}$ is varied, and in (b) as $\delta$ is varied. The results are averages from {100} runs. Error bars show one standard error (in both plots very small).
}%
\label{fig:fig_binary}%
\end{figure}

For the problem of determining the sign of $p_{1} - \frac{1}{2}$ from samples, the predominant approach in the literature is to construct lower and upper confidence bounds on $p_{1}$ that hold with probability $1 - \delta_{t}$ for each $t \geq 1$, satisfying $\sum_{t = 1}^{\infty} \delta_{t} \leq \delta$. The $\delta$-correctness of the procedure is ensured by  terminating only when the lower confidence bound exceeds $\frac{1}{2}$, or the upper confidence bound falls below $\frac{1}{2}$. We compare PPR-Bernoulli with several variants from the literature. In Figure~\ref{fig:fig_binary}, we plot the sample complexity of different algorithms as $p_{1}$ and  $\delta$ are varied.

A common choice is to set $\delta_{t} = k \frac{\delta}{t^{\alpha}}$, with constants $k$ and $\alpha$ tuned for efficiency, while ensuring $\delta$-correctness. As representatives of this approach, we pick the LUCB and KL-LUCB algorithms~\citep{pmlr-v30-Kaufmann13}. The former inverts Hoeffding's inequality to obtain lower and upper confidence bounds, while the latter uses a tighter Chernoff bound. Although these algorithms themselves are meant for bandit applications, their efficiency crucially depends on the tightness of the confidence bounds applied to each arm. The tuned confidence bounds~\citep{pmlr-v30-Kaufmann13} hence become suitable baselines for our comparison.\footnote{Details of all our implementations are given in Appendix~\ref{app:implementationdetails}; links to code are provided in Appendix~\ref{app:code}.}

With the intent of avoiding a na\"{i}ve union bound over time, \citet{garivier2013} applies a so-called \textit{peeling} argument to divide time into increasingly-sized slices. He obtains confidence regions by associating the random stopping time with a self-normalised process. The resulting stopping rule, which we denote KL-SN, still has a tunable parameter ``c'', which we set as recommended by \citet{garivier2013}. Although the $\mathcal{A}_{1}$ algorithm \citep{Shah_Choudhury_Karamchandani_Gopalan_2020} is designed specifically for mode estimation, we include it in this comparison to observe its performance when $K = 2$. In this special case, the algorithm reduces to an application of an empirical Bernstein bound~\citep{maurer2009}.



The two plots in Figure~\ref{fig:fig_binary} are remarkably consistent as $p_{1}$ and $\delta$ are varied. KL-LUCB shows a marginal improvement over (Hoeffding) LUCB, while KL-SN clearly outperforms both. However, PPR-Bernoulli is significantly more efficient than even KL-SN. Surprisingly, in spite of using variance information, $\mathcal{A}_{1}$ registers the worst performance among all the methods compared. We attribute this result to slack in the constants used in its stopping rule.


The empirical evidence of its sample efficiency, along with
its simplicity and non-reliance on parameter-tuning, make PPR-Bernoulli an attractive proposition for stopping problems. In Section~\ref{sec:ppr-me}, we consider three separate ways to generalise it to mode estimation. In Section \ref{sec:theory} we follow with theoretical analysis to explain the empirical findings in sections \ref{sec:pprmcs} and \ref{sec:ppr-me}.




%% file: sections/ppr-me.tex
\section{GENERALISATION TO \texorpdfstring{$K \geq 2$}{}}
\label{sec:ppr-me}

In the broader machine learning literature, the most common approaches for generalising $2$-class problems to more classes are ``one-versus-one'' (denoted 1v1) and ``one-versus-rest'' (denoted 1vr). We investigate both approaches. We also consider the direct application of the multi-dimensional (MD) variant of the PPR martingale test~\citep{waudby-ramdas-ppr}. 


\subsection{One-versus-one (1v1)  Approach}
\label{subsec:oneversusoneapproach}


In the first $t \geq 1$ samples, let the number of occurrences of value $v_{i}$ be $s^{t}_{i}$, $1 \leq i \leq K$. The 1v1 generalisation is based on the idea that if $v_{i}$ is to be declared the mode, we need to be sufficiently sure that $v_{i}$ is more probable than $v_{j}$ for $j \in \{1, 2, \dots, K\}, i \neq j$. 
Correspondingly, we simultaneously run PPR-Bernoulli tests on each $(i, j)$ pair with mistake probability $\frac{\delta}{K - 1}$. 
Each $(i, j)$ test relies solely on the number of occurrences of $v_{i}$ and $v_{j}$, disregarding other values. Hence it amounts to observing samples from a Bernoulli variable with parameter $\frac{p_{i}}{p_{i} + p_{j}}$, and verifying which side of $\frac{1}{2}$ its mean lies. The overall procedure stops when some $i \in \{1, 2, \dots, K\}$ has won each of its tests. By a union bound, with probability at least $1 - \delta$, the (true) mode $v_{1}$ will not ever lose a test. Thus, upon termination,  $v_{1}$ is returned with probability at least $1 - \delta$.

Whereas the description above suggests we need to monitor $\binom{K}{2}$ tests at each step, closer inspection reveals that a much lighter implementation is possible. As before, let $\text{first}(t)$ denote the index of the most-frequently occurring value (with arbitrary tie-breaking) after $t$ samples: that is, $s^{t}_{\text{first}(t)} \geq s^{t}_{i}$ for $i \in \{1, 2, \dots, K\}$. Now, if at all a winner is identified after $t$ samples, clearly it must be $v_{\text{first}(t)}$, which has as many occurrences as any other value. Hence, we only need to track tests involving $v_{\text{first}(t)}$. Now, it is also immediate that $v_{\text{first}(t)}$ wins all its tests if and only if it defeats the \text{second} most frequently occurring value, which we denote $\text{second}(t)$: that is, 
$\text{second}(t) \in \{1, 2, \dots, K\}, \text{second}(t) \neq \text{first}(t)$ satisfies $s^{t}_{\text{second}(t)} \geq s^{t}_{i}$ for $i \in \{1, 2, \dots, K\} \setminus \{\text{first}(t)\}$. Hence, we may implement our stopping rule, denoted PPR-1v1, using a \textit{single} PPR-Bernoulli test at each $t \geq 1$.
\begin{center}
\fbox{
\parbox{0.94\columnwidth}{
\textbf{PPR-1v1}: 
Stop and declare $v_{\text{first}(t)}$ as mode iff
$\phantom{aaaa}\text{Beta}\left(\frac{1}{2}; s^{t}_{\text{first}(t)} + 1, s^{t}_{\text{second}(t)} + 1\right) \leq \frac{\delta}{K - 1}$.
}
}
\end{center}

Tracking  $\text{first}(t)$ and
$\text{second}(t)$ is a simple computation; as observed earlier, it is also efficient to compute the Beta density at $\frac{1}{2}$.  Indeed our experiments show that PPR-1v1 is much faster computationally than other mode estimation algorithms (see Appendix~\ref{app:code}).

\subsection{One-versus-rest (1vr) Approach}
\label{subsec:oneversusrestapproach}


Notice that under PPR-1v1, sample $x^{t}$ at each step $t \geq 1$ contributes only to the $K - 1$ PPR-Bernoulli tests of the particular $i$ from $\{1, 2, \dots, K\}$ that satisfies $v_{i} = x^{t}$. The $\binom{K - 1}{2}$ tests corresponding to values other than $x^{t}$ receive no information. The 1vr approach becomes an alternative to address this \textit{apparent} wastage of information. Under the 1vr scheme, we associate a Bernoulli variable $B_{i}$ with each value $v_{i}$, $1 \leq i \leq K$, which has probability $p_{i}$ of generating $v_{i}$, and probability $1 - p_{i}$ of generating its \textit{negation} ``$\neg v_{i}$''. Consequently, each sample of $\mathcal{P}$ adds to one of the outcomes of $B_{i}$ for \textit{each} $i \in \{1, 2, \dots, K\}$. We draw an anytime confidence sequence for $B_{i}$ with mistake probability $\frac{\delta}{K}$, and terminate after $t \geq 1$ samples if the confidence set of $B_{\text{first}(t)}$ does not overlap with any of the others. Invoking the PPR martingale confidence sequence, we note that 
with probability at least $1 - \frac{\delta}{K}$, $p_{i}$ will lie  in all the intervals $(\text{LCB}^{t}_{i}, \text{UCB}^{t}_{i})$, $t \geq 1$, where $\text{LCB}^{t}_{i} = \min \{q \in [0, 1]: \pi^{t}(q) = \frac{\delta}{K} \}$ and $\text{UCB}^{t}_{i} = \max \{q \in [0, 1]: \pi^{t}(q) = \frac{\delta}{K} \}$
can be computed numerically. The $\delta$-correctness of the 1vr rule, given below, follows from a union bound on the mistake probabilities of each $B_{i}$, $1 \leq i \leq K$.
\begin{center}
\fbox{
\parbox{0.9\columnwidth}{
\textbf{PPR-1vr}: 
Stop and declare $v_{\text{first}(t)}$ as mode iff\\\phantom{a}for $1 \leq i \leq K$, $i \neq \text{first}(t)$, 
$\text{LCB}^{t}_{\text{first}(t)} \geq \text{UCB}^{t}_{i}.$
}
}
\end{center}
\subsection{Multi-dimensional (MD) PPR Test}
\label{subsec:ppr-md}

PPR Martingale confidence sequences can be directly constructed for the multi-dimensional parameter vector of $\mathcal{P}$~\citep[see Appendix C]{waudby-ramdas-ppr}. In this approach, denoted PPR-MD, at each $t \geq 1$ we maintain a confidence set ${C}^t$ with $\bar{p} \in [0,1]^{k}$, such  that $C^{t} \eqdef \{\bar{p}: R^{t}(\bar{p}) < \frac{1}{\delta}\}$. We stop at time $t$ when all $\bar{p} \in C^t$ have the same unique mode. With the Dirichlet distribution being the conjugate prior of the categorical distribution, $R^{t}(\bar{p})$ has a convenient form if initialised with a uniform prior: $$R^t(\bar{p}) = \frac{1}{(K-1)!}\times \frac{1}{\prod_{i=1}^{K}\bar{p}_i^{\alpha^{t}_i-1}} \times \frac{\prod_{i = 1}^{K} \Gamma(\alpha^{t}_i)}{\Gamma(\sum_{i = 1}^{K} \alpha^{t}_i)},$$ where for $1 \leq i \leq K$, $\alpha^{t}_{i} = s^{t}_{i} + 1$. Observe that this formulation reduces to PPR-Bernoulli for $K = 2$. However, for $K > 2$, checking for a unique mode in $C^{t}$ does not simplify to a convenient formula; it requires a numerical computation that increases steeply with $K$. We do not perform extensive experiments with PPR-MD---a choice justified by Lemma~\ref{lem:1vrimplies1v1} (Section~\ref{sec:theory}).

\subsection{Empirical Comparisons}
\label{subsec:empiricalcomparisonmodeestimation}

We compare PPR-1v1 and PPR-1vr with other mode estimation algorithms on a variety of discrete distributions. Table~\ref{tab:tab_multiK} summarises the results.

\begin{table*}[t]
\footnotesize
    \centering
    \scshape
\caption{Sample complexity comparison for mode estimation, run with mistake probability $\delta = 0.01$. The number after the $\times$ symbol indicates the multiplicity of that particular probability value in the distribution; thus $\mathcal{P}_{1}$ has $p = (0.5, 0.25, 0.25)$. The values reported are averages from 100 or more runs, and show one standard error.}
        \label{tab:tab_multiK}
    \begin{tabular}{l c c c c c}
    \hline
    \toprule
        Distribution & K & Type & $\mathcal{A}_{1}$ \citep{Shah_Choudhury_Karamchandani_Gopalan_2020} & KL-SN \citep{garivier2013}  & PPR \\
        \midrule
        \multirow{2}{*}{$\mathcal{P}_1$: .5, .25 $\times$ 2} & \multirow{2}{*}{3} & 1vr & 1344$\pm$20 & 418$\pm$14 & 262$\pm$12\\
         & & 1v1 & 1158$\pm$19 & 346$\pm$13 & \textbf{218$\pm$11} \\
         \midrule \multirow{2}{*}{$\mathcal{P}_2$: .4, .2 $\times$ 3} & \multirow{2}{*}{4} & 1vr & 1919$\pm$29 & 632$\pm$18 & 397$\pm$15\\
         & & 1v1 & 1516$\pm$24 & 468$\pm$15 & \textbf{298$\pm$13}\\
         \midrule \multirow{2}{*}{$\mathcal{P}_3$: .2, .1 $\times$ 8} & \multirow{2}{*}{9} & 1vr & 5082$\pm$51 & 1900$\pm$42 & 1201$\pm$29\\
         & & 1v1 & 3340$\pm$43 & 1138$\pm$31 & \textbf{789$\pm$28}\\
         \midrule \multirow{2}{*}{$\mathcal{P}_4$: {.1, .05 $\times$ 18 }} & \multirow{2}{*}{19} & 1vr & 12015$\pm$129 & 4686$\pm$81 & 2850$\pm$55\\
         & & 1v1 & 7352$\pm$88 & 2554$\pm$57 & \textbf{1840$\pm$53}\\
         \midrule \multirow{2}{*}{$\mathcal{P}_5$: {.35, .33, .12, .1 $\times$ 2}} & \multirow{2}{*}{5} & 1vr & 155277$\pm$2356 & 63739$\pm$2238 & 38001$\pm$1311\\
         & & 1v1 & 117988$\pm$2078 & 47205$\pm$1291 & \textbf{33660$\pm$1125}\\
         \midrule 
         
         
         \multirow{2}{*}{$\mathcal{P}_6$: {.35, .33,  .04 $\times$ 8}} & \multirow{2}{*}{10} & 1vr & 158254$\pm$2442 & 66939$\pm$2241 & 41963$\pm$1330\\
         & & 1v1 & 121150$\pm$2183 & 49576$\pm$1341 & \textbf{36693$\pm$1185}\\
        \bottomrule
        \end{tabular}
\normalsize
    \end{table*}

The $\mathcal{A}_{1}$ algorithm~\citep{Shah_Choudhury_Karamchandani_Gopalan_2020} is essentially a 1vr approach that uses Empirical Bernstein confidence bounds~\citep{maurer2009}. Noting that it can just as well be implemented in a 1v1 form, we include such a variant, denoted $\mathcal{A}_{1}$-1v1, in our experimental comparisons with PPR. For good measure, we also include 1v1 and 1vr variants based on the KL-SN confidence bound~\citep{garivier2013}, which finished second to PPR-Bernoulli for $K = 2$ (see Section~\ref{sec:pprmcs}). 

In Table~\ref{tab:tab_multiK}, we observe the same trend on \textit{each} problem instance: (1) The 1v1 variant of each stopping rule outperforms the corresponding 1vr variant, and (2) PPR is most sample-efficient, followed by KL-SN and $\mathcal{A}_{1}$. Although the prohibitive running time of PPR-MD prevents a thorough assessment, a few informal runs indicate that its sample complexity is well in excess of even PPR-1vr.

\begin{figure}[b!]
\centering
\includegraphics[width=\columnwidth]{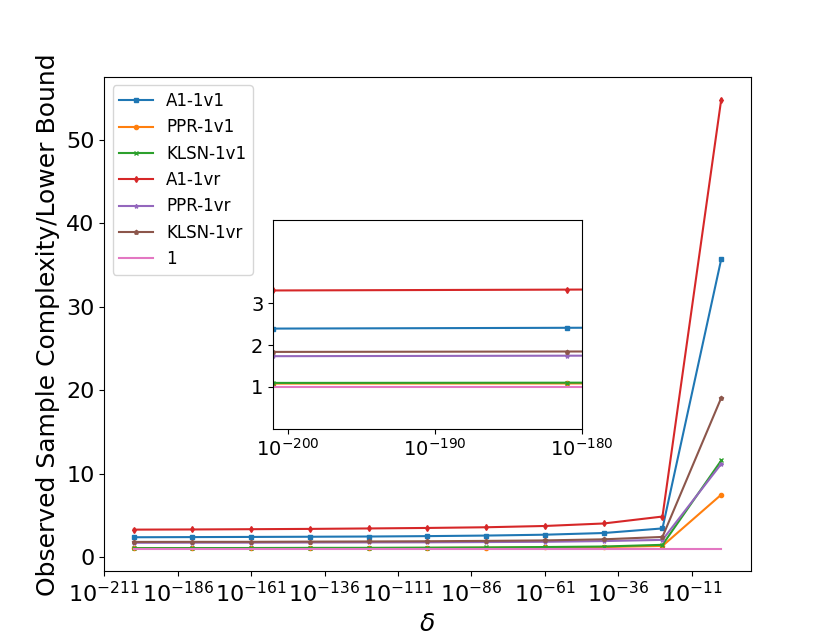}
\caption{Comparison of different stopping rules on $\mathcal{P}_{3}$, for small values of $\delta$. The y axis plots the ratio of the empirical stopping time (averaged over 100 or more runs) and $\text{LB}(\mathcal{P}_{3}, \delta)$, defined in Theorem~\ref{thm:lowerbound}.}
\label{fig:asymptotic}
\end{figure}

Whereas the results in  Table~\ref{tab:tab_multiK} are for a fixed value of $\delta = 0.01$, we conduct a second set of experiments to compare the performance of the different algorithms as $\delta$ is varied. In particular, we investigate the ``asymptotic'' regime, in which $\delta$ is taken to $0$. In Figure~\ref{fig:asymptotic}, we plot ratio of the empirical sample complexity and the lower bound from Theorem~\ref{thm:lowerbound}, varying $\delta$ while keeping the distribution fixed to $\mathcal{P}_{3}$ from Table~\ref{tab:tab_multiK}. Observe that once again, the relative order among the algorithms remains the same. The 1v1 variant of each algorithm performs better than its 1vr counterpart. Moreover, the curves for
PPR-1v1 and KLSN-1v1 suggest that these rules might be asymptotically optimal.

The empirical evidence supporting the PPR martingale test and the 1v1 approach to mode estimation is compelling. In the forthcoming section, we provide theoretical reasons to explain our observations.\footnote{After this paper was submitted for review, the authors were pointed to recent related work by \citet{haddenhorst}, who focus on the identification of a generalised Condorcet winner in multi-dueling bandits. Mode estimation is a special case of the problem they consider. Their algorithm, based on the Dvoretzky-Kiefer-Wolfowitz (DKW) inequality, has an upper bound which improves upon that given by \citet{Shah_Choudhury_Karamchandani_Gopalan_2020}.
In particular, the bound is independent of $K$, and depends on $\frac{1}{(p_1-p_2)^2}\ln\ln\frac{1}{(p_1-p_2)}$, rather than $\frac{1}{(p_1-p_2)^2}\ln\frac{1}{(p_1-p_2)}$. However, our experiments show that the two variants of their algorithm---DKW-1, which is provided the knowledge of $p_{1} - p_{2}$, and DKW-2, which has no such prior knowledge---both perform worse than PPR-1v1. For example,  when run on problem instance with $K = 2, p_1 = 0.66, \delta=0.01$, PPR-Bernoulli requires roughly 124 samples, whereas DKW-1 and DKW-2 take roughly 469 and 871 samples, respectively. The inferior empirical performance of the latter algorithms in spite of their superior upper bound can be explained by the accompanying constant factors. The notion of asymptotic optimality that we present in sections \ref{sec:ppr-me} and \ref{sec:theory} requires even the constant factor to be tight, albeit as $\delta \to 0$.}

%% file: sections/theory.tex
\section{THEORETICAL JUSTIFICATION}
\label{sec:theory}

In our experiments, we observe that not only do the 1v1 variants of each method perform better than 1vr in aggregate, they terminate \textit{before} the 1vr variants on \textit{every single run}. We prove this result true for some of the methods. We also formally establish that PPR-MD and $\mathcal{A}_{1}$ cannot terminate before PPR-1v1.
\begin{lemma}
\label{lem:1vrimplies1v1}
Let $X = x^{1}, x^{2}, \dots$ be an infinite sequence of samples from $\mathcal{P}$. For algorithm $\mathcal{L}$ and $\delta \in (0, 1)$, let $T(\mathcal{L}, X, \delta)$ be the stopping time of $\mathcal{L}$ on $X$, when run with mistake probability $\delta$. For algorithms $\mathcal{L}_{1}, \mathcal{L}_{2}$, let the proposition $\mathcal{G}(\mathcal{L}_{1}, \mathcal{L}_{2}, X, \delta)$ denote ``If $T(\mathcal{L}_{2}, X, \delta)$ is finite, then $T(\mathcal{L}_{1}, X, \delta) \leq T(\mathcal{L}_{2}, X, \delta)$''. For all $\mathcal{P}$, for all $X$ generated from $\mathcal{P}$, for all $\delta \in (0, 1)$, we have
\begin{itemize}
\item[(i)] $\mathcal{G}(\text{LUCB-1v1}, \text{LUCB-1vr}, X, \delta)$,
\item[(ii)] $\mathcal{G}(\mathcal{A}_{1}\text{-1v1}, \mathcal{A}_{1}\text{-1vr}, X, \delta)$,
\item[(iii)] $\mathcal{G}(\text{PPR-1v1}, \text{PPR-MD}, X, \delta)$,
\item[(iv)] $\mathcal{G}(\text{PPR-1v1}, \mathcal{A}_{1}\text{-1v1}, X, \delta)$.
\end{itemize}
\end{lemma}

We give a proof of the lemma in Appendix~\ref{app:relatingpprmeterminationandpprevrtermination}. The proofs of (i) and (ii) formalise the intuition that although the 1vr variants update all their $K$ tests with each sample, the test to separate any two variables is less effective than that under the corresponding 1v1 variant. We conjecture that $\mathcal{G}(\text{PPR-1v1}, \text{PPR-1vr}, X, \delta)$ is also true, in fact verifying it to be the case for $|X| \leq 200$. We also believe that a similar result applies to the 1v1 and 1vr variants of KL-LUCB and KL-SN (but the proofs become cumbersome). Result (iii) is straightforward to establish from first principles. To show (iv), consider that after $t$ samples, $\mathcal{A}_{1}$-1v1 draws upper and lower confidence bounds on the mean as $\hat{p}^{t} \pm \beta^{t}$, where $\hat{p}^{t}$ is the empirical mean and $\beta^{t}$ the confidence width. In particular, $\beta^t = \beta^{t}_{1} + \beta^{t}_{2}$, where $\beta^{t}_{1} = \sqrt{\frac{2V^t\text{ln}(4 t^2/\alpha)}{t}}$ and $\beta^{t}_{2} = \frac{7\text{ln}(4 t^2/\alpha)}{3(t-1)}$, with $V^t$ being the empirical variance and $\alpha$ the input mistake probability. When $\hat{p}^{t}$ is within a constant distance from $0.5$ (hence $V^{t}$ is close to its maximum of $0.25$), we show that the PPR confidence set is contained in $[\hat{p}^{t} - \beta^{t}_{1}, \hat{p}^{t} + \beta^{t}_{1}]$. When $\hat{p}^{t}$ is either sufficiently small or sufficiently large, $t$ (at termination) can itself be upper-bounded in terms of $\hat{p}^{t}$, in turn lower-bounding $\beta^{t}_{2}$ and guaranteeing that the PPR confidence set is contained in $[\hat{p}^{t} - \beta^{t}_{2}, \hat{p}^{t} + \beta^{t}_{2}]$.

Lemma~\ref{lem:1vrimplies1v1} offers theoretical justification for many of the trends observed in Table~\ref{tab:tab_multiK}. The theoretical question that remains open is an explanation of Figure~\ref{fig:asymptotic}: is PPR-1v1 indeed asymptotically optimal? We obtain an affirmative answer.


\begin{theorem}[Optimality of PPR-1v1]
\label{thm:asymptotic}
Fix $\delta \in (0, 1)$, $K \geq 2$, and distribution $\mathcal{P} = (p, v, K)$. Let $\tau(\mathcal{P}, \delta)$ be the expected stopping time of PPR-1v1 on $\mathcal{P}$ when run with mistake probability $\delta$, and let $\text{LB}(\mathcal{P}, \delta)$ be the lower bound defined in Theorem~\ref{thm:lowerbound}. Then $$\lim_{\delta \to 0} \frac{\tau(\mathcal{P}, \delta)}{\text{LB}(\mathcal{P}, \delta)} = 1.$$
\end{theorem}

The proof of the theorem is given in Appendix~\ref{app:optimality-final}. 
In the proof, we establish that with a little over $\text{LB}(\mathcal{P}, \delta)$ samples from $\mathcal{P}$, a sufficient number of samples are obtained for separating value $v_{1}$ from each of the others. The proof of separation for each pair uses a similar sequence of steps as of \citet{pmlr-v49-garivier16a}.




In short, our theoretical analysis reinforces PPR-1v1 as the method of choice for PAC mode estimation. For good measure, Appendix~\ref{app:ppr-bernoulli-upperboundproof} presents an explicit upper bound on the sample complexity of PPR-1v1, which holds for all $\delta \in (0, 1)$. This bound improves upon that of $\mathcal{A}_{1}$~\citep{Shah_Choudhury_Karamchandani_Gopalan_2020} by a constant factor.


%% file: sections/applications.tex
\section{PRACTICAL APPLICATIONS}
\label{sec:practicalapplications}

Our main motivation for devising better mode estimation algorithms is their practical significance, which we illustrate through two contrasting applications.


\subsection{Forecasting in Indirect Elections}
\label{sec:elections}

Opinion polls 
to forecast the winner of an upcoming election
are a natural application of mode estimation. In fact, the algorithms discussed in Section~\ref{sec:ppr-me} can all be applied with only minor alterations to  \textit{plurality} systems, wherein the task is precisely that of determining the choice preferred by the largest fraction of the target population. \citet{waudby-ramdas-ppr} illustrate the use of the PPR martingale test 
on this application, while focusing on  \textit{without-replacement} sampling. In parliamentary democracies such as India~\citep{rajeeva2002} and the U.K.~\citep{payne2003}, a two-level voting system is used to elect governments. In this system,  individuals in each \textit{constituency} (or \textit{seat})---typically a geographically contiguous region---elect a party based on plurality; the party winning the most seats forms the government. 
Forecasting the winning party in such an \textit{indirect} voting system calls for a more sophisticated sampling procedure. Whereas it would \textit{suffice} to separately identify the winner from each seat by sampling, it might be wasteful to do so when the overall winning party has a clear majority in its number of seats. 

Formally, consider a setting where we have $K$ parties and $N$ constituencies. Each constituency $c \in \{1, 2, \dots, N\}$ represents a discrete probability distribution $\mathcal{P}^c=(p^c, v, K)$, where for $i \in \{1, 2, \dots, K\}$, $v_i$ represents the political party $i$ and $p^c_i$ denotes the fraction of votes won by party $i$ in constituency $c$ (for simplicity we have assumed all parties compete in all constituencies). If $i^{\star}_{c}$ is the index of the mode of $\mathcal{P}^{c}$ (assumed unique), our objective is to determine
$$\argmax_{i \in \{1, 2, \dots, K\}} \sum_{c = 1}^{N} \mathbf{1}[i = i^{\star}_{c}]$$ correctly with probability at least $1 - \delta$. The objective of the sampling rule is to minimise the total votes sampled, $\sum\limits_{c=1}^{N}T^c$, where $T^c$ represents the number of votes queried in  constituency $c$.


We consider a procedure that 
(1) keeps track of the current winners and leaders at the aggregate level, and (2) at each step samples the constituencies that appear most promising to confirm the aggregate trend. In principle, this algorithm, denoted DCB (for ``Difference in Confidence Bounds'') can be coupled with any algorithm that uses confidence bounds for mode estimation. Yet, we obtain the best results when DCB uses PPR-1v1 as a subroutine, thereby highlighting the relevance of PPR-1v1 not only as a stopping rule, but also as an input to on-line decision making.

DCB takes cue from the LUCB algorithm for best-arm identification in bandits~\citep{ICML12-shivaram}. At each step $t$, it identifies two parties, $a^{t}$ and $b^{t}$, that appear the most promising to win the overall election: these parties are picked based on their current number of wins and ``leads'' in individual constituencies. Subsequently the algorithm chooses a constituency each for $a^{t}$ and $b^{t}$, samples from which could ``most'' help distinguish the tally of the two. We provide a detailed specification of DCB in Appendix~\ref{app:dcbalgorithm}.

\begin{table}[t]
\footnotesize
\centering
\scshape
  \caption{Sample complexity of various stopping rules when coupled with (1) round-robin (RR) polling of constituencies and (2) DCB. All experiments are run with mistake probability $\delta=0.01$. Values shown are averages from 10 runs, and show one standard error.  ``Seats resolved'' indicates the number of constituencies in which a winner was identified before the overall procedure terminated.}
  \label{tab:elections-india}  \begin{tabular}{lccccc}
    
    \toprule
    \multirow{3}{*}{Algorithm} & 
      \multicolumn{2}{c}{India-2014 (543 seats)} &
      \\
      & \multirow{2}{*}{Samples} & {Seats}\\
      &  & {Resolved} \\
      \midrule
    RR-$\mathcal{A}_1$-1v1 & 1578946 $\pm$ 10255 & 239 $\pm$ 2 \\
    RR-$\mathcal{A}_1$-1vr & 1767464 $\pm$ 15828 & 234 $\pm$ 1 \\
    RR-KLSN-1v1 & 610181 $\pm$ 7072 & 240 $\pm$ 2 \\
    RR-KLSN-1vr & 726512 $\pm$ 6186 & 236 $\pm$ 2 \\
    RR-PPR-1v1 & \textbf{471661 $\pm$ 7373} & 241 $\pm$ 3 \\
    RR-PPR-1vr & 560815 $\pm$ 7778 & 238 $\pm$ 2  \\\midrule
    DCB-$\mathcal{A}_1$-1v1 & 856678 $\pm$ 3935 & 182 $\pm$ 1 \\
    DCB-$\mathcal{A}_1$-1vr & 951684 $\pm$ 5246 & 180 $\pm$ 1 \\
    DCB-KLSN-1v1 & 325265 $\pm$ 2096 & 186 $\pm$ 2 \\
    DCB-KLSN-1vr & 376108 $\pm$ 4067 & 181 $\pm$ 1 \\
    DCB-PPR-1v1 & \textbf{256911 $\pm$ 2096} & 188 $\pm$ 1 \\
    DCB-PPR-1vr & 296580 $\pm$ 2372 & 184 $\pm$ 1 \\
    \bottomrule
  \end{tabular}
\normalsize
\end{table}





We compare DCB with a round-robin strategy for picking the next constituency to sample. Both approaches can be implemented with different stopping rules, which are also varied. Table~\ref{tab:elections-india} shows our results on the 2014 parliamentary elections in India.\footnote{Election results are in the public domain; the authors accessed them at \url{https://www.indiavotes.com/}.} In this election, the winning party secured 282 seats from among 543, giving it a very large victory over the second-place party, which won 44 seats.
Results from a much closer contest, in the state of Bihar, are given in Appendix~\ref{app:bihar-table}. We see a similar trend in both cases.


While it is not the central feature of this paper, it is worth noting that the DCB strategy indeed improves over round-robin polling by roughly a factor of two, regardless of the stopping rule. As intended, it does not waste samples on constituencies that are inconsequential to the overall result (observed in the ``seats resolved'' columns). Of more direct relevance to the theme of the paper is that even when embedded within a decision-making outer loop, PPR continues to outperform $\mathcal{A}_{1}$ and KL-SN, and the 1v1 approach still dominates 1vr. Surmising that the sheer efficiency of PPR-1v1 makes it a good choice for embedding in more complex systems, we proceed to our next application.

\subsection{Verifying Blockchain Smart Contracts}
\label{sec:blockhains}

Our second application of PAC mode estimation is in a domain of growing contemporary relevance. Permissionless blockchains such as Bitcoin~\citep{NakamotoBitcoin} and Ethereum~\citep{buterin2014} 
allow uncertified agents to join a pool of service providers, also called \textit{nodes}.
A recent feature that has emerged in such blockchains is the execution of ``smart contracts''~\citep{sourav2019,buterin2014}, which could include, for example, running computationally-heavy jobs such as machine learning algorithms. In an ideal world, a client who requires some computation performed can simply enter a smart contract with some particular node, and pay a fee for the service. Unfortunately, there is no guarantee that nodes in a permissionless blockchain are honest. A ``Byzantine''
(\textit{malicious}) node could potentially return a quick-to-compute, incorrect output, to the detriment of the client. 

In recent work, \citet{sourav2019} propose an approach for the \textit{probabilistic verification} of smart contracts. Abstractly, assume that the computation to be performed for the client is deterministic, and it has a (yet unknown) output $o_{\text{correct}}$. The proposed model accommodates any blockchain in which the fraction of Byzantine nodes $f$ is at most $f_{\max} \in [0, \frac{1}{2})$. With this assumption, it becomes feasible to give a probabilistic guarantee on obtaining the correct output. For any fixed mistake probability $\delta \in (0, 1)$, the client could ship out the computation to  $\Theta\left(\frac{1}{(\frac{1}{2} - f_{\max})^{2}}\log(\frac{1}{\delta})\right)$
nodes, and take their majority response as the answer, thereby ensuring $\delta$-correctness. Unfortunately, transaction costs can be substantial, especially those for computationally-intensive contracts. Hence, it is in the client's interest to minimise the number of nodes queried to achieve the same probabilistic guarantee. For example, a sequential procedure could potentially query fewer nodes if $f \ll f_{\max}$.


Our contribution in the context of this application is to propose PPR-1v1 as an alternative to the Sequential Probability Ratio Test (SPRT)~\citep{awald1945}, which is used by \citet{sourav2019} for their verification procedure. This classical test finds use in many other engineering applications~\citep{kenny1991,chen2008}, some of which could also benefit from the advantages of PPR-1v1 over SPRT.

\begin{figure}[!ht]
    \centering
    \subfloat[\centering $K  = 2$ (single adversarial answer)]{{\includegraphics[width=\columnwidth]{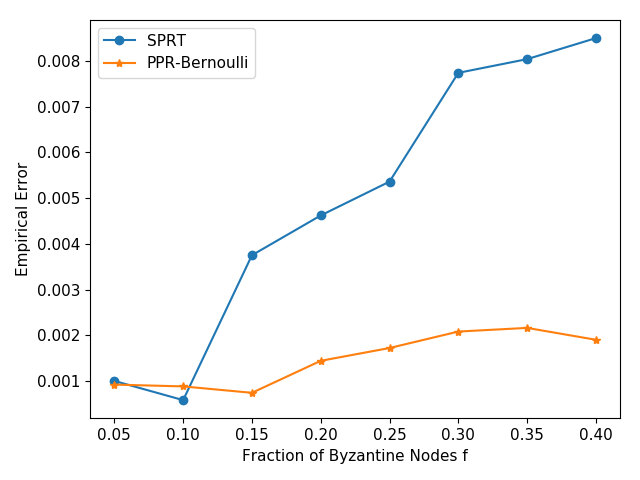} 
    }
    \label{fig:fmax-underestimate}
    }%
    
    \subfloat[\centering $K = 10$  ]{{\includegraphics[width=\columnwidth]{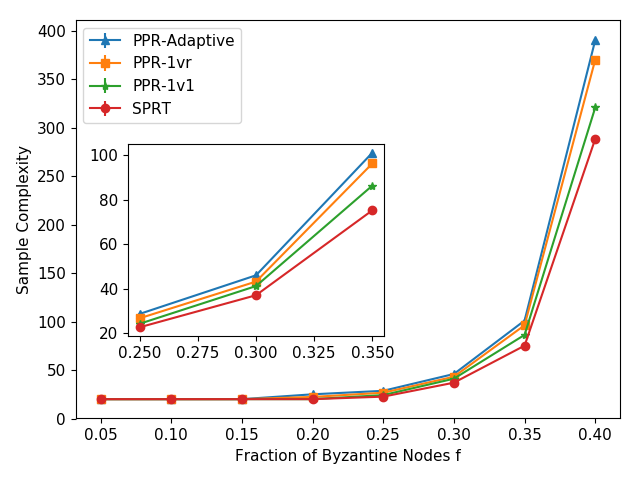}
    }
    \label{fig:adaptiveK}
    }%
    \caption{Comparisons with SPRT for the probabilistic verification of smart contracts, obtained with parameter settings
    $N = 1600$, $m = 20$, $\delta = 0.005$, $f_{max} = 0.1$. Plot (a) shows the empirical error rates of the algorithms as the true Byzantine fraction $f$ is varied, taking $K = 2$. Plot (b) shows the sample complexity of various algorithms against varying $f$, on an instance with $K = 10$ answers.}%
    \label{fig:blockchain}%
\end{figure}
To apply SPRT for verifying smart contracts, \citet{sourav2019} assume that out of the total of $N$ nodes in the blockchain, batches of size $m$, chosen uniformly at random, can be queried in sequence (the $m$ queries per batch are performed \textit{in parallel}, hence saving some time).
For simplicity assume the answers returned are from the set $\{0, 1, 2, \dots\}$. Let $q = \frac{m}{N}$, and let $c_{i,t}$ be the number of times answer $i$ is reported in the $t^{th}$ step, $i \geq 0$, $t \geq 1$. Defining $l_{i,T} = \sum_{t=1}^{T} (2c_{i,t} - m)m$, a derivation~\citep{sourav2019} establishes that $\delta$-correct SPRT stops at time $T$, giving $i$ as the answer if
\begin{equation*}
    l_{i,T} > \ln\left ( \frac{1 - \delta}{\delta} \right) \frac{2q(1-q)N (1 - f_{\max})f_{\max}}{1 - 2f_{\max}}.
\end{equation*}
The primary disadvantage of SPRT herein is the need for the user to provide $f_{\max}$, which is used in the stopping rule.
While a lower value of $f_{\max}$ will improve the efficiency of the rule, unfortunately $\delta$-correctness no longer holds if $f$, the true fraction of Byzantine nodes, exceeds $f_{max}$. Figure~\ref{fig:fmax-underestimate} plots the empirical error made by SPRT (averaged over 50,000 runs) on a problem instance in which the Byzantine nodes all give the same (incorrect) answer. We fix $f_{\max} = 0.1$, and plot the error by varying $f$. The blockchain has $N = 1600$ nodes, of which SPRT samples $m = 20$ at a time. Although the test is run using  mistake probability $\delta = 0.005$, observe that the empirical error exceeds $\delta$ when $f > f_{\max}$.
Since the verification task at hand is precisely that of PAC mode estimation, PPR-1v1 becomes a viable alternative, especially since it does not need the knowledge of $f_{\max}$. In fact, PPR-1v1 can identify the mode even if its associated probability is less than $\frac{1}{2}$ (although in this case, it can no longer be guaranteed that the mode is $o_{\text{correct}}$, since the Byzantine nodes may collude). Observe from Figure~\ref{fig:fmax-underestimate} that unlike SPRT, the empirical error rate of PPR-1v1 (equivalent to PPR-Bernoulli since we have set $K = 2$) remains within $\delta$ even for $f > f_{\max}$.

In Figure~\ref{fig:adaptiveK}, we compare the sample complexities of PPR-1v1 and SPRT. Whereas other problem parameters (including $f_{max}$) stay the same as before, we consider an instance in which $K = 10$.
The single correct answer is given by a $(1 - f)$-fraction of the nodes, while $9$ different incorrect answers are given by the Byzantine nodes, each equally common. The version of SPRT used is a 1vr
adaptation of the basic procedure~\citep{sourav2019} to $K = 10$. First, we observe that SPRT terminates before PPR-1v1 and PPR-1vr at all values of $f$. The PPR algorithms pay this price for having to assure $\delta$-correctness at all values of $f < \frac{1}{2}$, unlike SPRT, which does so only for $f < f_{max}$. In the same plot,
we show the performance of another PPR variant, denoted ``PPR-Adaptive''. In reality, we cannot be sure about the number of answers $K$ that will be returned by the blockchain's nodes---and hence cannot use it in our stopping rule. Under PPR-Adaptive, which is a 1v1 strategy, the overall mistake probability $\delta$ is divided into the \textit{infinite} sequence $k \frac{\delta}{1^{2}}, k \frac{\delta}{2^{2}}, k \frac{\delta}{3^{2}}, \dots$ (with $k = \frac{6}{\pi^{2}}$). Whenever a new answer is revealed, it is inserted into the list of possible answers, and its pairwise tests given mistake probabilities from the unused portion of the sequence. In principle, PPR-Adaptive can accommodate any number of answers, incurring only a small increase in sample complexity, as visible from Figure~\ref{fig:adaptiveK}.

%% file: sections/conclusion.tex
\section{CONCLUSION}
\label{sec:conclusion}

In this paper, we apply the framework of PPR Martingale confidence sequences to the problem of PAC mode estimation. Our investigation follows two different dimensions that play a significant role in determining the efficiency of stopping rules. First is the tightness of the confidence bounds used internally in the stopping rule. By separately focusing on the Bernoulli case, we show that the PPR Martingale stopping rule is sample-efficient. The second aspect of mode estimation is the template applied to generalise from $K = 2$ to $K \geq 2$, which can typically be applied with any valid confidence bounds. Of the three major choices---``one-versus-one'' (1v1),  ``one-versus-rest'' (1vr), and a multi-dimensional test (MD)---we find  1v1 to be the most efficient. Our empirical findings are affirmed by theoretical analysis that shows (1) regardless of the problem instance and the mistake probability, the 1v1 approach is guaranteed to terminate no later than the 1vr approach for many popular confidence bounds, and (2) the PPR-1v1 stopping rule is indeed asymptotically optimal. The PPR-1v1 algorithm is also parameter-free and computationally much faster than the other algorithms, making it a natural choice to apply to practical mode estimation problems. We illustrate its efficacy on two distinct real-world tasks.

Our paper opens several directions to explore in future work, including the application of the PPR martingale test on pure exploration problems in stochastic bandits and Markov Decision Problems. It could also be of much practical benefit to incorporate the PPR martingale test (in place of existing ones) in large-scale applications of sampling and decision making.

%% file: sections/acknowledgements.tex
\section*{Acknowledgements}
\label{sec:acks}

The authors thank anonymous reviewers for providing valuable suggestions. Shivaram Kalyanakrishnan was   partially supported   by SERB grant ECR/2017/002479.

%% file: appendices/implementationdetails.tex
\section{IMPLEMENTATION DETAILS}

\label{app:implementationdetails}
We specify the various baseline algorithms used in our comparisons in sections
\ref{sec:pprmcs} and
\ref{sec:ppr-me}. For actual code see Appendix~\ref{app:code}.

Whether the implementation is 1v1 or 1vr, each algorithm has an atomic operation to maintain upper and lower confidence bounds on the parameter of a Bernoulli distribution.
\begin{itemize}
\item In the 1v1 approach, there is a separate Bernoulli associated with each pair of values $v_{i}$ and $v_{j}$ for $1 \leq i < j \leq k$ (hence a Bernoulli with mean $\frac{p_{i}}{p_{i} + p_{j}}$). Each confidence bound is drawn with mistake probability $\frac{\delta}{K - 1}$, which ensures $\delta$-correctness upon termination.

\item In the 1vr approach, there is a separate Bernoulli associated with each value $v_{i}$ for $i \leq i \leq k$, taking $\neg v_{i}$ as its other outcome (hence a Bernoulli with mean $p_{i}$). To ensure $\delta$-correctness upon termination, the mistake probability used is $\frac{\delta}{K}$.
\end{itemize}

The actual confidence bounds used in different algorithms are listed below.
In 1v1, the total number of samples ($t$) shown in the confidence bounds is to be taken as the sum of the number of occurrences of the corresponding $(v_{i}, v_{j})$ pair in question, while under 1vr, it is the total number of samples of $\mathcal{P}$ observed yet. Below we denote the empirical mean after $t$ samples $\hat{p}^{t}$, and the mistake probability associated with the confidence bound $\delta$.

\subsection{\texorpdfstring{$\mathcal{A}_{1}$~\citep{Shah_Choudhury_Karamchandani_Gopalan_2020}}{}}
We implement the algorithm as given in the original paper~\citep{Shah_Choudhury_Karamchandani_Gopalan_2020}, which defines a confidence width
\begin{equation*}
    \beta(t, \delta) = \sqrt{\frac{2V^t\text{ln}(4 t^2/\delta)}{t}} + \frac{7\text{ln}(4 t^2/\delta)}{3(t-1)}
\end{equation*}
after $t$ samples, where $V^t$ is the empirical variance of the samples. Lower and upper confidence bounds are given by $\hat{p}^{t} \pm \beta(t, \delta)$.


\subsection{LUCB \texorpdfstring{\citep{ICML12-shivaram}}{} and KL-LUCB \texorpdfstring{\citep{pmlr-v30-Kaufmann13}}{}}
For LUCB~\citep{ICML12-shivaram} and KL-LUCB~\citep{pmlr-v30-Kaufmann13}, the ``exploration rate'' used is $\beta(t, \delta) = \text{ln}(\frac{405.5 t^{1.1}}{\delta})$, which ensures $\delta$ correctness when a union bound over $t$ is performed~\citep{pmlr-v30-Kaufmann13}. 

LUCB sets its lower and upper confidence bounds as $\hat{p}^{t} \pm \sqrt{\frac{\beta(t, \delta)}{2t}}$. KL-LUCB obtains them by performing a numerical computation to obtain the lower and upper confidence bounds as given below.


\begin{align*}
\text{KL-LUCB lower confidence bound } &= \min \{q \in [0, \hat{p}^{t}]: t \times D_{KL}(\hat{p}^{t}||q) \leq \beta(t, \delta) \};\\
\text{KL-LUCB upper confidence bound } &= \max \{q \in [\hat{p}^{t}, 1]: t \times D_{KL}(\hat{p}^{t}||q) \leq \beta(t, \delta) \}.
\end{align*}

\subsection{KL-SN \texorpdfstring{\citep{garivier2013}}{}}
KL-SN~\citep{garivier2013} uses a more sophisticated exploration rate so as to avoid a na{\"i}ve union bound over time. We first find  $\gamma > 1$ that satisfies the following equation, and thereafter set the exploration rate as given below.
\begin{equation*}
    2e^2\gamma e^{-\gamma} = \delta \text{ and } \beta(t, \delta) = \frac{\gamma (1 + \text{ln}(\gamma))}{(\gamma - 1)\text{ln}(\gamma)} \text{ln(ln}(t)) + \gamma.
\end{equation*}

Using the above exploration rate, the confidence bounds are constructed in the same way as KL-LUCB.

\begin{align*}
\text{KL-SN lower confidence bound } &= \min \{q \in [0, \hat{p}^{t}]: t \times D_{KL}(\hat{p}^{t}||q) \leq \beta(t, \delta) \}; \\
\text{KL-SN upper confidence bound } &= \max \{q \in [\hat{p}^{t}, 1]: t \times D_{KL}(\hat{p}^{t}||q) \leq \beta(t, \delta) \}.
\end{align*}

%% file: appendices/code.tex
\section{CODE DETAILS}
\label{app:code}

The code used to run our experiments (from sections \ref{sec:pprmcs}, \ref{sec:ppr-me}, \ref{sec:elections}, and \ref{sec:blockhains}) is at \url{https://github.com/rohanshah13/pac_mode_estimation}. Below we provide details on the running time of our algorithms on a couple of problem instances, which are indicative of their relative order in general.

The results in Table~\ref{tab:tab_multiK} were obtained by performing each of the experiments for $100$ iterations, with mistake probability $\delta = 0.01$. For the last $3$ rows, for all algorithms except KL-SN 1v1, $200$ iterations were used so as to reduce the error bars. A running-time comparison of all the algorithms across the $100$ runs is given in Table \ref{tab:multiK_runtime}. The runs were performed on an Intel Core i7-8750H CPU @ 2.20GHz processor.

\begin{table}[H]
\footnotesize
\centering
\scshape
  \caption{Average running time and one standard error of mode estimation algorithms across $100$ iterations, distribution $\mathcal{P}_6 = (.35, .33, .04\times8)$ from Table~\ref{tab:tab_multiK}.}
  \label{tab:multiK_runtime} . \begin{tabular}{lcc}
    \toprule
    Algorithm & Type & Run time (seconds)\\
      \midrule
      \multirow{2}{*}{$\mathcal{A}_1$} & 1vr & 9.01 $\pm$ 0.2 \\
      & 1v1 & 6.65 $\pm$ 0.17 \\
      \midrule
      \multirow{2}{*}{KL-SN} & 1vr & 169.47 $\pm$ 5.76 \\
      & 1v1 & 2.74 $\pm$ 0.10 \\
      \midrule
      \multirow{2}{*}{PPR} & 1vr & 51.5 $\pm$ 2.37 \\
      & 1v1 & \textbf{1.28 $\pm$ 0.06} \\
    \bottomrule
  \end{tabular}
 \vspace{0.1cm}
\normalsize
\end{table}

The results in Table~\ref{tab:elections-india} and Table~\ref{tab:elections-bihar} were obtained by performing each of the experiments for 10 random seeds and setting the mistake probability to $0.01$. The sampling of the votes from each of the constituencies were done with a batch size of 200. We used Google's colaboratory services~\footnote{\url{https://colab.research.google.com/}} for performing these experiments. 

%% file: appendices/1v1and1vrtermination.tex
\newpage
\section{RELATING THE TERMINATION OF 1V1, 1VR, MD ALGORITHMS} \label{app:relatingpprmeterminationandpprevrtermination}
In this section, we provide a proof of Lemma~\ref{lem:1vrimplies1v1} and also comment on the plausible applicability of the result to concentration bounds not covered by the lemma. Since the actual working is relatively verbose, we divide the appendix into separate subsections. In Section~\ref{app:1v1-1vr-lucb}, we provide a relatively straightforward proof that the 1v1 versions of LUCB and $\mathcal{A}_1$ always terminate before their corresponding 1vr versions. In Section~\ref{app:1v1-1vr-ppr}, we work towards proving that PPR-1v1 always terminates before PPR-1vr, and we reduce this claim to an inequality on beta functions, which we conjecture to be true, and have verified empirically for a range of values. In Section~\ref{app:1v1-1vr-pprmd}, we consider the multi-dimensional version of the PPR martingale stopping rule (denoted PPR-MD), which serves as an alternative for designing a stopping rule for mode estimation. This rule 
would use a Dirichlet prior (which is the conjugate of the multinomial distribution), and is analogous to the mutlti-variate PPR considered in Appendix C of \cite{waudby-ramdas-ppr}. We show that PPR-MD always stops after PPR-1v1 on every run. In Section \ref{app:ppr-a1}, we provide a simple proof that shows $\mathcal{A}$1-1v1 \citep{Shah_Choudhury_Karamchandani_Gopalan_2020} always terminates after PPR-1v1 on every run.

In summary, sections \ref{app:1v1-1vr-lucb}, \ref{app:1v1-1vr-pprmd} and \ref{app:ppr-a1} complete the proof of Lemma~\ref{lem:1vrimplies1v1}, while Section~\ref{app:1v1-1vr-ppr} concludes with a conjecture, which if true, would mean the termination of PPR-1vr implies the termination of PPR-1v1.

\subsection{LUCB and \texorpdfstring{$\mathcal{A}_1$}{} Algorithms}
\label{app:1v1-1vr-lucb}

\textbf{LUCB: } Suppose we run both the variants of LUCB algorithm with $\delta' = \frac{\delta}{K}$, where $\delta$ is the mistake probability (in reality, we run the 1v1 with $\frac{\delta}{K-1}$; so it is even better). We will show that LUCB-1vr termination implies LUCB-1v1 termination. The LUCB algorithm can differentiate some $i,j \in \{1,2,..,K\}$ such that $i \neq j$ when
\begin{gather*}
    \hat{p}_i^t - \sqrt{\frac{\ln\left(\frac{405.5t^{1.1}}{\delta'}\right)}{2t}} \geq \hat{p}_j^t + \sqrt{\frac{\ln\left(\frac{405.5t^{1.1}}{\delta'}\right)}{2t}}, 
\end{gather*}
where $\hat{p}_i^t$ denotes the empirical mean corresponds to observation $i$. For 1v1, the corresponding condition for differentiating $i, j$ will be
\begin{gather*}
    \frac{\hat{p}_i^t}{\hat{p}_i^t + \hat{p}_j^t} - \sqrt{\frac{\ln\left(\frac{405.5(s_{ij}^t)^{1.1}}{\delta'}\right)}{2s_{ij}^t}} \geq \frac{\hat{p}_j^t}{\hat{p}_i^t + \hat{p}_j^t} + \sqrt{\frac{\ln\left(\frac{405.5(s_{ij}^t)^{1.1}}{\delta'}\right)}{2s_{ij}^t}},
\end{gather*}
where $s_{ij}^t$ denotes the total number of samples coming from $p_i$ and $p_j$. Thus, to prove 1vr termination implies 1v1 termination, It is sufficient to prove that: If it is true that
\begin{gather*}
    \hat{p}_i^t - \sqrt{\frac{\ln\left(\frac{405.5t^{1.1}}{\delta'}\right)}{2t}} \geq \hat{p}_j^t + \sqrt{\frac{\ln\left(\frac{405.5t^{1.1}}{\delta'}\right)}{2t}} 
\end{gather*}
then, it is also true that
\begin{gather*}
    \frac{\hat{p}_i^t}{\hat{p}_i + \hat{p}_j^t} - \sqrt{\frac{\ln\left(\frac{405.5(s_{ij}^t)^{1.1}}{\delta'}\right)}{2s_{ij}^t}} \geq \frac{\hat{p}_j^t}{\hat{p}_i^t + \hat{p}_j} + \sqrt{\frac{\ln(\frac{405.5(s_{ij}^t)^{1.1}}{\delta'})}{2s_{ij}^t}}.
\end{gather*}
We have
\begin{gather*}
    \hat{p}_i^t - \hat{p}_j^t \geq 2\sqrt{\frac{\ln\left(\frac{405.5t^{1.1}}{\delta'}\right)}{2t}}, \text{ }  \\
    \implies \frac{\hat{p}_i^t - \hat{p}_j^t}{2(\hat{p}_i^t + \hat{p}_j^t)\sqrt{\frac{\ln\left(\frac{405.5(s_{ij}^t)^{1.1}}{\delta'}\right)}{2s_{ij}^t}}} \geq \frac{\sqrt{\frac{\ln\left(\frac{405.5t^{1.1}}{\delta'}\right)}{2t}}}{(\hat{p}_i^t + \hat{p}_j^t)\sqrt{\frac{\ln\left(\frac{405.5(s_{ij}^t)^{1.1}}{\delta'}\right)}{2s_{ij}^t}}}.
\end{gather*}
Using the fact that $\hat{p}_i^t + \hat{p}_j^t = \frac{s_{ij}^t}{t}$ and the above inequality, we get
\begin{gather*}
    \frac{\hat{p}_i^t - \hat{p}_j^t}{2(\hat{p}_i^t + \hat{p}_j^t)\sqrt{\frac{\ln\left(\frac{405.5(s_{ij}^t)^{1.1}}{\delta'}\right)}{2s_{ij}^t}}} \geq \frac{\sqrt{t\ln\left(\frac{405.5t^{1.1}}{\delta'}\right)}}{\sqrt{s_{ij}^t\ln\left(\frac{405.5(s_{ij}^t)^{1.1}}{\delta'}\right)}}.  
\end{gather*}
Since $\sqrt{t\ln\left(\frac{405.5t^{1.1}}{\delta'}\right)}$ is an increasing function in $t$, and we know that $s_{ij}^t \leq t$, we have
\begin{gather*}
    \frac{\hat{p}_i^t - \hat{p}_j^t}{2(\hat{p}_i^t + \hat{p}_j^t)\sqrt{\frac{\ln\left(\frac{405.5(s_{ij}^t)^{1.1}}{\delta'}\right)}{2s_{ij}^t}}} \geq \frac{\sqrt{t\ln\left(\frac{405.5t^{1.1}}{\delta'}\right)}}{\sqrt{s_{ij}^t\ln(\frac{405.5(s_{ij}^t)^{1.1}}{\delta'})}} \geq 1,\\
    \implies \frac{\hat{p}_i^t}{\hat{p}_i^t + \hat{p}_j^t} - \sqrt{\frac{\ln\left(\frac{405.5(s_{ij}^t)^{1.1}}{\delta'}\right)}{2s_{ij}^t}} \geq \frac{\hat{p}_j^t}{\hat{p}_i^t + \hat{p}_j^t} + \sqrt{\frac{\ln\left(\frac{405.5(s_{ij}^t)^{1.1}}{\delta'}\right)}{2s_{ij}^t}}.
\end{gather*}
Hence, we showed that the 1v1 variant of LUCB algorithm always terminates before its 1vr variant. \\\\
\textbf{$\mathcal{A}_1$ Algorithm:} Suppose we run both the variants of $\mathcal{A}_1$ algorithm with $\delta' = \frac{\delta}{K}$, where $\delta$ is the mistake probability (As above, we actually run 1v1 with $\frac{\delta}{K-1}$, so 1v1 will terminate even faster). The $\mathcal{A}_1$ algorithm differentiates between $i,j \in \{1,2,..,K\}$ such that $i \neq j$ and $i$ is the winner when
\begin{gather*}
    \hat{p}_i^t - \sqrt{\frac{2V_t(Z^i) \ln \left(\frac{4t^2}{\delta'}\right)}{t}} - \frac{7 \ln\left( \frac{4t^2}{\delta'} \right) }{3(t-1)} \geq \hat{p}_j^t + \sqrt{\frac{2V_t(Z^j) \ln \left(\frac{4t^2}{\delta'}\right)}{t}} + \frac{7 \ln\left( \frac{4t^2}{\delta'} \right) }{3(t-1)},
\end{gather*}
where
\begin{gather*}
    V_t (Z^i) = \frac{s_i^t(t - s_i^t)}{t(t-1)},
\end{gather*}
in which $s_i^t$ denotes the total number of samples coming from $v_i$. Similarly, the condition for there to be a winner between $i,j$ in $\mathcal{A}_1$ 1v1 will be:
\begin{gather*}
    \frac{\hat{p}_i^t}{\hat{p}_i^t + \hat{p}_j^t} - \sqrt{\frac{2V_{s_{ij}^t}(Z^i) \ln \left(\frac{4(s_{ij}^t)^2}{\delta'}\right)}{s_{ij}^t}} - \frac{7 \ln\left( \frac{4(s_{ij}^t)^2}{\delta'} \right) }{3(s_{ij}^t-1)} \geq \frac{\hat{p}_j^t}{\hat{p}_i^t + \hat{p}_j^t} + \sqrt{\frac{2V_{s_{ij}^t}(Z^j) \ln \left(\frac{4(s_{ij}^t)^2}{\delta'}\right)}{s_{ij}^t}} + \frac{7 \ln\left( \frac{4(s_{ij}^t)^2}{\delta'} \right) }{3(s_{ij}^t-1)},
\end{gather*}
where $s_{ij}^t = s_i^t + s_j^t$ denotes the total number of samples coming from $v_i$ and $v_j$, and $s_i^t$ denotes the number of samples coming from $v_i$.\\
To prove 1vr termination implies 1v1 termination, It is sufficient to prove that: If it is true that
\begin{gather*}
     \hat{p}_i^t - \sqrt{\frac{2V_{t}(Z^i) \ln \left(\frac{4t^2}{\delta'}\right)}{t}} - \frac{7 \ln\left( \frac{4t^2}{\delta'} \right) }{3(t-1)} \geq  \hat{p}_j^t + \sqrt{\frac{2V_{t}(Z^j) \ln \left(\frac{4t^2}{\delta'}\right)}{t}} + \frac{7 \ln\left( \frac{4t^2}{\delta'} \right) }{3(t-1)} 
\end{gather*}
then, it is also true that
\begin{gather*}
     \frac{\hat{p}_i^t}{\hat{p}_i^t + \hat{p}_j^t} - \sqrt{\frac{2V_{s_{ij}^t}(Z^i) \ln \left(\frac{4(s_{ij}^t)^2}{\delta'}\right)}{s_{ij}^t}} - \frac{7 \ln\left( \frac{4(s_{ij}^t)^2}{\delta'} \right) }{3(s_{ij}^t-1)} \geq \frac{\hat{p}_j^t}{\hat{p}_i^t + \hat{p}_j^t} + \sqrt{\frac{2V_{s_{ij}^t}(Z^j) \ln \left(\frac{4(s_{ij}^t)^2}{\delta'}\right)}{s_{ij}^t}} + \frac{7 \ln\left( \frac{4(s_{ij}^t)^2}{\delta'} \right) }{3(s_{ij}^t-1)} 
\end{gather*}
Using the same steps as in LUCB, we get
\begin{gather*}
    \frac{\hat{p}_i^t - \hat{p}_j^t}{2(\hat{p}_i^t + \hat{p}_j^t)\left(\sqrt{\frac{2V_{s_{ij}^t}(Z^i) \ln \left(\frac{4(s_{ij}^t)^2}{\delta'}\right)}{s_{ij}^t}} + \sqrt{\frac{2V_{s_{ij}^t}(Z^j) \ln \left(\frac{4(s_{ij}^t)^2}{\delta'}\right)}{s_{ij}^t}} + \frac{14 \ln\left( \frac{4(s_{ij}^t)^2}{\delta'} \right) }{3(s_{ij}^t-1)}  \right)} \\ \geq  \frac{\left(\sqrt{2t V_{t}(Z^i) \ln \left(\frac{4t^2}{\delta'}\right)} + \sqrt{2t V_{t}(Z^j) \ln \left(\frac{4t^2}{\delta'}\right)} + \frac{14t \ln\left( \frac{4t^2}{\delta'} \right) }{3(t-1)}  \right)}{\left(\sqrt{2s_{ij}^t V_{s_{ij}^t}(Z^i) \ln \left(\frac{4(s_{ij}^t)^2}{\delta'}\right)} + \sqrt{2s_{ij}^t V_{s_{ij}^t}(Z^j) \ln \left(\frac{4(s_{ij}^t)^2}{\delta'}\right)} + \frac{14s_{ij}^t \ln\left( \frac{4(s_{ij}^t)^2}{\delta'} \right) }{3(s_{ij}^t-1)}  \right)}. 
\end{gather*}
We need to prove that the above expression is $\geq 1$. We note that $\sqrt{2t V_{t}(Z^i) \ln \left(\frac{4t^2}{\delta'}\right)}$, $\sqrt{2t V_{t}(Z^j) \ln \left(\frac{4t^2}{\delta'}\right)}$ are increasing functions in $t$. Let $f(t) = \frac{14t \ln\left( \frac{4t^2}{\delta'} \right) }{3(t-1)}$. We write down the differential:
\begin{gather*}
    f'(t) \propto 2(t-1)-\ln\left(\frac{4t^2}{\delta}\right).
\end{gather*}
Thus, for $t$ such that $2(t-1) \geq \ln\left(\frac{4t^2}{\delta}\right)$, $f(t)$ is an increasing function. We now look at the minimum value of $s_i^t$ possible, and we show that $s_i^t$ is always such that $f(t)$ is an increasing function for $t \geq s_i^t$. We proceed to look at a scenario in which lowest possible value for $s_i^t$ is observed. Consider the case in which we observe $t\hat{p}_i^t$ samples with value $v_i$ (thus, the empirical mode is $\hat{p}_i^t$ for $v_i$). Let's assume that $\mathcal{A}_1$ 1vr algorithm declares $i$ as the mode after observing $t\hat{p}_i^t$ samples of $v_i$. We assume that $j$ has $0$ samples, and then find the value of $t$ that arises. Termination of $\mathcal{A}_1$ algorithm after $t$ observations suggests (we ignore the empirical mean term, as it makes our lower bound on $t$ only bigger):

\begin{gather*}
    \hat{p}_i^t - \frac{7\ln\left(\frac{4t^2}{\delta'}\right)}{3(t-1)} \geq 0 + \frac{7\ln\left(\frac{4t^2}{\delta'}\right)}{3(t-1)} \\
    \implies \hat{p}_i^t \geq \frac{14\ln\left(\frac{4t^2}{\delta'}\right)}{3(t-1)}.
\end{gather*}

Using the fact that $t\hat{p}_i^t = s_i^t$, we arrive at:
\begin{gather*}
    \frac{3\hat{p}_i^t(t-1)}{14} \geq \ln\left(\frac{4t^2}{\delta'}\right)\\
    \implies \frac{3(s_i^t - \hat{p}_i^t)}{14} \geq \ln\left(\frac{4t^2}{\delta'}\right) \geq \ln\left(\frac{4(s_i^t)^2}{\delta'}\right).
\end{gather*}

To show $2(t_i-1) \geq \ln\left(\frac{4(s_i^t)^2}{\delta}\right)$, it is sufficient to show that $2(s_i^t-1) \geq \frac{3}{14}(s_i^t - \hat{p}_i^t)$. For proving this, we make a mild assumption that $s_i^t \geq 2$ as \emph{practically} $\mathcal{A}_1$ algorithm would need more than 2 samples from $v_i$ to declare $v_i$ as the mode. Under this assumption, we get:
\begin{gather*}
    2(s_i^t-1) - \frac{3}{14}(s_i^t - \hat{p}_i^t) = \frac{25}{14}s_i^t - 2 + \frac{3}{14}\hat{p}_i^t \geq \frac{11}{7} + \frac{3}{14}\hat{p}_i^t > 0.
\end{gather*}

Hence, we note that 
\begin{gather*}
    \frac{\left(\sqrt{2t V_{t}(Z^i) \ln \left(\frac{4t^2}{\delta'}\right)} + \sqrt{2t V_{t}(Z^j) \ln \left(\frac{4t^2}{\delta'}\right)} + \frac{14t \ln\left( \frac{4t^2}{\delta'} \right) }{3(t-1)}  \right)}{\left(\sqrt{2s_{ij}^t V_{s_{ij}^t}(Z^i) \ln \left(\frac{4(s_{ij}^t)^2}{\delta'}\right)} + \sqrt{2s_{ij}^t V_{s_{ij}^t}(Z^j) \ln \left(\frac{4(s_{ij}^t)^2}{\delta'}\right)} + \frac{14s_{ij}^t \ln\left( \frac{4(s_{ij}^t)^2}{\delta'} \right) }{3(s_{ij}^t-1)}  \right)} \geq 1
\end{gather*}
since all the $3$ terms are increasing, and thus $\mathcal{A}_1$ 1vr always stops after $\mathcal{A}_1$ 1v1.

\subsection{PPR}
\label{app:1v1-1vr-ppr}

Suppose party $i, j$ have confidence sequences which do not intersect at some time $t$. Let the number of samples from party $i$ be $s_i^t$, and the number of samples from party $j$ be $s_j^t$, and without loss of generality assume $s_i^t > s_j^t$. Let $f$ be the number of samples from parties which are not $i$ or $j$. Then, the condition for party $i, j$ having confidence sequences which do not intersect can be written as 
\begin{gather*}
    \exists \: \theta \in \left[\frac{s_j^t}{t}, \frac{s_i^t}{t} \right] \text{ such that } \\
    \frac{1}{\text{Beta}\left(\theta; s_i^t+1,t-s_i^t+1\right)} \geq \frac{K}{\delta}, \\
    \frac{1}{\text{Beta}\left(\theta; s_j^t+1,t-s_j^t+1\right)} \geq \frac{K}{\delta}.
\end{gather*}
Parties $i, j$ have a winner between them in PPR-1v1 when
\begin{gather*}
    \frac{1}{\text{Beta}\left(\frac{1}{2};s_i^t+1,s_j^t+1\right)} \geq \frac{K-1}{\delta}.
\end{gather*}
We will attempt to show that party $i, j$ having disjoint confidence sequences in PPR-1vr implies party $i,j$ having a winner between them in PPR-1v1. This will imply that PPR-1v1 always terminates before PPR-1vr. Noting that
\begin{gather*}
    \text{Beta}\left(\theta;s_i^t,s_j^t\right) = \frac{\theta^{s_i^t-1}(1-\theta)^{s_j^t-1}}{B(s_i^t,s_j^t)} \\
    \text{where } B(s_i^t,s_j^t) = \frac{\Gamma(s_i^t)\Gamma(s_j^t)}{\Gamma(s_i^t+s_j^t)},
\end{gather*}
we are given that 
\begin{gather*}
    \frac{\delta}{K} \geq \frac{\theta^{s_i^t}(1-\theta)^{t-s_i^t}}{B(s_i^t+1,t-s_i^t+1)}, \\
    \frac{\delta}{K} \geq \frac{\theta^{s_j^t}(1-\theta)^{t-s_j^t}}{B(s_j^t+1,t-s_j^t+1)}.
\end{gather*}
And we wish to show that 
\begin{gather*}
    \frac{\delta}{K-1} \geq \frac{1}{2^{s_i^t+s_j^t}B(s_i^t+1,s_j^t+1)}.
\end{gather*}
We denote
\begin{gather*}
    \frac{\theta^{s_i^t}(1-\theta)^{t-s_i^t}}{B(s_i^t+1,t-s_i^t+1)} = L_1(\theta), \\
    \frac{\theta^{s_j^t}(1-\theta)^{t-s_j^t}}{B(s_j^t+1,t-s_j^t+1)} = L_2(\theta). 
\end{gather*}

Thus, it is sufficient to show that, $\forall \: \theta \in \left[\frac{s_j^t}{t}, \frac{s_i^t}{t} \right]$,
\begin{gather*}
    \max\left(L_1(\theta), L_2(\theta)\right) \geq \frac{K-1}{K2^{s_i^t+s_j^t}B(s_i^t+1,s_j^t+1)}.
\end{gather*}
Consider
\begin{gather*}
    L_1'(\theta) \propto s_i^t(1-\theta) - (t-s_i^t)\theta, \\
    L_1'(\theta) \propto s_i^t - t\theta \\
    \implies L_1'(\theta) \geq 0 \: \forall \: \theta \leq \frac{s_i^t}{t}.
\end{gather*}
Similarly, consider 
\begin{gather*}
    L_2'(\theta) \propto s_j^t(1-\theta) - (t-s_j^t)\theta \\
    L_2'(\theta) \propto s_j^t - t\theta, \\
    \implies L_2'(\theta) \leq 0 \: \forall \: \theta \geq \frac{s_j^t}{t}.
\end{gather*}
Thus, we note that in our range of $\theta, L_1(\theta)$ is increasing and $L_2(\theta)$ is decreasing. Consider $\theta^*$ such that $L_1(\theta^*) = L_2(\theta^*)$. We will prove that considering only the value $\theta = \theta^*$ is sufficient to prove that 
\begin{gather*}
    \max\left(L_1(\theta), L_2(\theta)\right) \geq \frac{K-1}{K2^{s_i^t+s_j^t}B(s_i^t+1,s_j^t+1)}
\end{gather*}
for the entire range of $\theta \in \left[\frac{s_j^t}{t}, \frac{s_i^t}{t} \right]$. First, we prove that $\theta^*$ lies in this range itself. The equation $L_1(\theta^*) = L_2(\theta^*)$ gives us the implicit equation
\begin{gather*}
    \left(\frac{\theta^*}{1-\theta^*}\right) = \left[\frac{s_i^t!(t-s_i^t)!}{s_j^t!(t-s_j^t)!}\right]^{\frac{1}{s_i^t-s_j^t}}.
\end{gather*}
We first prove $\theta^* \geq \frac{s_j^t}{t}$. This is the same as showing that $\frac{\theta^*}{1-\theta^*} \geq \frac{s_j^t}{t-s_j^t}$. We show that 
\begin{gather*}
    \frac{s_j^t}{t-s_j^t} \leq \left[\frac{s_i^t!(t-s_i^t)!}{s_j^t!(t-s_j^t)!}\right]^{\frac{1}{s_i^t-s_j^t}} \\ 
    \Leftrightarrow \left(\frac{s_j^t}{t-s_j^t}\right)^{s_i^t-s_j^t} \leq \left[\frac{(s_j^t+1)(s_j^t+2)...(t-s_i^t)}{(s_i^t+1)(s_i^t+2)...(t-s_j^t)}\right].
\end{gather*}
To prove the above inequality, first note that 
\begin{gather*}
    \frac{s_j^t}{t-s_j^t} \leq \frac{s_j^t+1}{t-s_j^t}.
\end{gather*}
Hence, it is sufficient now to show
\begin{gather*}
    \left(\frac{s_j^t}{t-s_j^t}\right)^{s_i^t-s_j^t-1} \leq \left[\frac{(s_j^t+2)...(t-s_i^t)}{(s_i^t+1)(s_i^t+2)...(t-s_j^t-1)}\right].
\end{gather*}
We know that $s_j^t < s_i^t$. Suppose that $s_j^t = s_i^t-1$; then the above equation has both LHS = 1 and RHS = 1, hence it is true. Otherwise, if $s_j^t < s_i^t-1$, we have 
\begin{gather*}
    \frac{s_j^t}{t-s_j^t} \leq \frac{s_j^t+2}{t-s_j^t-1}.
\end{gather*}
And proceeding in a similar fashion, we can show inductively that 
\begin{gather*}
    \frac{s_j^t}{t-s_j^t} \leq \left[\frac{s_i^t!(t-s_i^t)!}{s_j^t!(t-s_j^t)!}\right]^{\frac{1}{s_i^t-s_j^t}}.
\end{gather*}
In the same way, we can also show that
\begin{gather*}
    \frac{s_i^t}{t-s_i^t} \geq \left[\frac{s_i^t!(t-s_i^t)!}{s_j^t!(t-s_j^t)!}\right]^{\frac{1}{s_i^t-s_j^t}}.
\end{gather*}
This proves that $\theta^* \in \left[\frac{s_j^t}{t}, \frac{s_i^t}{t} \right]$. Moreover, since $L_1(\theta)$ is increasing and $L_2(\theta)$ is decreasing, we have that 
\begin{gather*}
    L_1(\theta^*) \leq L_1(\theta), \theta \geq \theta^*, \\
    L_2(\theta^*) \geq L_2(\theta), \theta \leq \theta^*. \\
\end{gather*}
Combining the above two equations and noting that $L_1(\theta^*) = L_2(\theta^*)$, we get that 
\begin{gather*}
    \max\left(L_1(\theta^*), L_2(\theta^*)\right) \leq \max\left(L_1(\theta), L_2(\theta)\right), \theta \in \left[\frac{s_j^t}{t}, \frac{s_i^t}{t} \right].
\end{gather*}
Thus, if we show that 
\begin{gather*}
    \max\left(L_1(\theta^*), L_2(\theta^*)\right) \geq \frac{K-1}{K2^{s_i^t+s_j^t}B(s_i^t+1,s_j^t+1)}
\end{gather*}
we are done. We need to show that 
\begin{gather*}
    \frac{\left(\theta^{*}\right)^{s_i^t}(1-\theta^*)^{t-s_i^t}}{B(s_i^t+1,t-s_i^t+1)} \geq \frac{K-1}{K2^{s_i^t+s_j^t}B(s_i^t+1,s_j^t+1)}, \\
    \text{where } \left(\frac{\theta^*}{1-\theta^*}\right) = \left[\frac{s_i^t!(t-s_i^t)!}{s_j^t!(t-s_j^t)!}\right]^{\frac{1}{s_i^t-s_j^t}}.
\end{gather*}
We ran computer simulations to verify the correctness of the above equation exhaustively for all values of $s_i^t, s_j^t, t$ in $[1, 200]$. Thus, we conjecture that the above equation is true for all $s_i^t, s_j^t, t > 0$ such that $s_i^t > s_j^t$. 

\subsection{Comparison with Multidimensional PPR (PPR-MD)}
\label{app:1v1-1vr-pprmd}

The Dirichlet distribution is the conjugate prior of the Multinomial distribution, and hence can be used to implement a PPR-based stopping rule, which is similar in principle to the multi-variate PPR given in Appendix C in \cite{waudby-ramdas-ppr}. In specific, the PDF of the Dirichlet distribution is given as 
\begin{gather*}
    pdf(\bar{p}) = \frac{\prod_{i=1}^{K}\bar{p}_i^{\alpha_i-1}}{B(\alpha)}, \\
    B(\alpha) = \frac{\prod_i \Gamma(\alpha_i)}{\Gamma(\sum_i \alpha_i)}.
\end{gather*}
At each time-step $t$, we maintain a confidence set ${C}^t$ with $\bar{p} \in [0,1]^{k}$, such  that $C^{t} \eqdef \{\bar{p}: R^{t}(\bar{p}) < \frac{1}{\delta}\}$. We stop at time $t$ when all the $\bar{p} \in C^t$ have the same unique mode. We call this stopping rule PPR-MD (Multi-dimensional). We note that, with prior $\alpha = [1, ..., 1]$,
\begin{gather*}
    R^t(\bar{p}) = \frac{1}{(K-1)!}\times \frac{B(\alpha)}{\prod_{i=1}^{K}\bar{p}_i^{\alpha_i-1}}
\end{gather*}
We will show that if the PPR-MD stopping rule stops at a point, then PPR-1v1 would also have stopped. In other words, PPR-MD stopping \textit{implies} PPR-1v1 stopping on every run.

Let the number of observations from $v_1, v_2, ..., v_K$ until timestep $t$ be $s_1^t, s_2^t, ..., s_K^t$. Without loss of generality, order them such that $s_1^t \geq s_2^t \geq ... \geq s_K^t$. We denote the empirical means as $\hat{p}_i^t = \frac{s_i^t}{t}$. We note that, when PPR-MD stops, $x_1 = (\frac{\hat{p}_1^t+\hat{p}_2^t}{2}, \frac{\hat{p}_1^t+\hat{p}_2^t}{2}, \hat{p}_3^t, ..., \hat{p}_K^t)$ must not lie in the confidence set. Explicitly,
\begin{gather*}
    \frac{\left(\frac{\hat{p}_1^t+\hat{p}_2^t}{2}\right)^{s_1^t+s_2^t}(\hat{p}_3^t)^{s_3^t}...\hat{p}_K^{s_K^t}}{s_1^t!s_2^t!...s_K^t!} \leq \frac{\delta(s_1^t+s_2^t+\ldots+s_K^t+K-1)!}{(K-1)!}.
\end{gather*}
We divide our proof into two parts. First, we show the implication for $K = 3$, and then we extend our proof to general $K \geq 3$.

\subsubsection{Proof for \texorpdfstring{$K = 3$}{}}
When $K = 3$, PPR-1v1 stops when
\begin{gather*}
    \frac{\left(\frac{1}{2}\right)^{s_1^t+s_2^t}(s_1^t+s_2^t+1)!}{s_1^t!s_2^t!} \leq \frac{\delta}{2}
\end{gather*}
and PPR-MD stopping implies that (using that $K = 3$)
\begin{gather*}
    \frac{\left(\frac{\hat{p}_1^t+\hat{p}_2^t}{2}\right)^{s_1^t+s_2^t}(\hat{p}_3^t)^{s_3^t}(s_1^t+s_2^t+s_3^t+2)!}{s_1^t!s_2^t!s_3^t!} \leq \frac{\delta}{2}.
\end{gather*}
If we can show that
\begin{gather*}
    \frac{\left(\frac{\hat{p}_1^t+\hat{p}_2^t}{2}\right)^{s_1^t+s_2^t}(\hat{p}_3^t)^{s_3^t}(s_1^t+s_2^t+s_3^t+2)!}{s_1^t!s_2^t!s_3^t!} \geq \frac{\left(\frac{1}{2}\right)^{s_1^t+s_2^t}(s_1^t+s_2^t+1)!}{s_1^t!s_2^t!},
\end{gather*}    
for all $s_1^t, s_2^t, s_3^t$ such that $s_1^t \geq s_2^t \geq s_3^t$, then PPR-MD stopping will imply PPR-1v1 stopping for $K = 3$.

 We note that this is equivalent to showing
\begin{gather*}
    (\hat{p}_1^t+\hat{p}_2^t)^{s_1^t+s_2^t}(\hat{p}_3^t)^{s_3^t} \geq \frac{s_3^t!(s_1^t+s_2^t+1)!}{(s_1^t+s_2^t+s_3^t+2)!} \\
    \Leftrightarrow \frac{(s_1^t+s_2^t)^{s_1^t+s_2^t}(s_3^t)^{s_3^t}}{(s_1^t+s_2^t+s_3^t)^{s_1^t+s_2^t+s_3^t}} \geq \frac{s_3^t!(s_1^t+s_2^t+1)!}{(s_1^t+s_2^t+s_3^t+2)!} \\
    \Leftrightarrow \frac{x^{x}y^{y}}{(x+y)^{x+y}} \geq \frac{(x+1)!y!}{(x+y+2)!}, \text{ where } x = s_1^t+s_2^t, y = s_3^t
\end{gather*}
for all $x, y,$ such that $y \geq 0, x \geq 2y$. Note that, when $y = 0$, the $y^y$ term tends to $1$; hence, the above statement is true. Subsequently, we will consider $y > 0$. Let 
\begin{gather*}
    f(x,y) = \frac{x^xy^y(x+2)....(x+y+2)}{y!(x+y)^{x+y}}
\end{gather*}
We want to show that 
\begin{gather*}
    f(x, y) \geq 1 \text{ when } y > 0, x \geq 2y.
\end{gather*}
We show this in two steps. First, we show that $f(x,y)$ is an increasing function of $x$ when $y$ is fixed. Since $f(x,y)$ is increasing in $x$, we then choose the minimum value of $x$ with fixed $y$ (i.e, $x = 2y$) and prove that $f(2y, y) \geq 1$ for all $y > 0$, which implies that $f(x, y) \geq 1$ for all $y > 0, x \geq 2y$.

\paragraph{Showing that $f(x,y)$ is Increasing in $x$}
\hfill\break
\hfill\break
Recapping, we need to show that 
\begin{gather*}
    f(x,y) = \frac{y^y(x+2)...(x+y+2)}{y!(x+y)^y\left(1+\frac{y}{x}\right)^x}
\end{gather*}
is increasing with respect to $x$ while keeping $y$ fixed. We write down the partial derivative of $\ln(f(x,y))$ with respect to $x$ as follows:
\begin{gather*}
    \frac{\partial \ln(f(x,y))}{\partial x} = \sum_{j=2}^{y}\frac{1}{x+j} - \frac{y}{x+y} + \frac{1}{x+y+2}+\frac{1}{x+y+1} + 1 - \frac{x}{x+y} - \ln\left(\frac{x+y}{x}\right) \\
    \implies \frac{\partial \ln(f(x,y))}{\partial x} = \sum_{j=2}^{y}\frac{1}{x+j} + \frac{1}{x+y+2}+\frac{1}{x+y+1} - \ln\left(\frac{x+y}{x}\right).
\end{gather*}
We note that
\begin{gather*}
    \frac{1}{x+j} \geq \int_{w=j}^{w=j+1}\frac{1}{x+w}dw = \ln\left(\frac{x+j+1}{x+j}\right) \\
    \implies \sum_{j=2}^{y}\frac{1}{x+j} \geq \ln\left(\frac{x+y+1}{x+2}\right).
\end{gather*}
Thus, if we show that 
\begin{gather*}
    g(x, y) = \ln\left(\frac{x+y+1}{x+2}\right) + \frac{1}{x+y+2}+\frac{1}{x+y+1} - \ln\left(\frac{x+y}{x}\right) \geq 0 \text{ when } y > 0, x \geq 2y
\end{gather*}
this is sufficient to show that $\frac{\partial \ln(f(x, y))}{\partial x} \geq 0 \text{ when } y \geq 0, x \geq 2y$. We combine terms to get
\begin{gather*}
    g(x, y) = \frac{1}{x+y+2}+\frac{1}{x+y+1} - \ln\left(\frac{(x+y)(x+2)}{(x+y+1)x}\right)
\end{gather*}
We will now show that keeping $x$ fixed, $g(x, y)$ is a decreasing function of $y$. Hence, it will be sufficient to consider the maximum possible value of $y$ to show that $g(x, y) \geq 0$.
We see that 
\begin{gather*}
    \frac{\partial g(x,y)}{\partial y} = -\left(\frac{3x^3+x^2(9y+11)+x(9y^2+22y+13)+3y^3+11y^2+13y+4}{(x+y)(x+y+1)^2(x+y+2)^2}\right) \\
    \implies \frac{\partial g(x,y)}{\partial y} \leq 0, y > 0, x \geq 2y. \\
\end{gather*}
Hence, $g(x, y)$ is decreasing in $y$. Thus, it is enough to consider the maximal possible value of $y$ for a fixed $x$ to show that $g(x,y) \geq 0$. For a fixed $x$, the maximum possible $y$ is $\frac{x}{2}$. Hence, we have that 
\begin{gather*}
    h(x) = g(x, x/2) = \frac{2}{3x+4} + \frac{2}{3x+2} - \ln\left(\frac{3(x+2)}{(3x+2)}\right),
\end{gather*}
and we need to show that $h(x) \geq 0 \text{ when } x > 0$. We have
\begin{gather*}
    h'(x) = -\left(\frac{8(9x^2+21x+14)}{(x+2)(3x+2)^2(3x+4)^2}\right) \\
    \implies h'(x) \leq 0 \text{ when } x > 0.
\end{gather*}
Thus, $h(x)$ is a decreasing function of $x$. We see that, as $x \rightarrow \infty, h(x) \rightarrow 0$. Thus, $h(x) \geq 0 \text{ when } x > 0$. This means that $g(x, x/2) \geq 0 \implies g(x, y) \geq 0 \implies \frac{\partial f(x,y)}{\partial x} \geq 0$ which implies that $f(x,y)$ is an increasing function of $x$.
\paragraph{Showing that $f(x,y) \geq 1$}
\hfill\break
\hfill\break
Since $f(x,y)$ is an increasing function of $x$ when $y$ is fixed, it is sufficient to show that $f(x,y) \geq 1$ for the minimum possible value of $x$. We note that the minimum possible value of $x$ is $x = 2y$. We have
\begin{gather*}
    f(2y, y)-1 = \frac{y^y(2y+2)...(3y+2)-y!(3y)^y\left(\frac{3}{2}\right)^{2y}}{y!(3y)^y\left(\frac{3}{2}\right)^{2y}}.
\end{gather*}
The denominator is positive. We show that the numerator is positive for all $y > 0$. 
\begin{gather*}
    h(2y, y) = y^y(2y+2)...(3y+2)-y!(3y)^y\left(\frac{3}{2}\right)^{2y} \\
    \implies h(2y, y) = y^y\left((2y+2)...(3y+2) - y!\left(\frac{27}{4}\right)^y \right).
\end{gather*}
Let 
\begin{gather*}
    c(y) = (2y+2)...(3y+2) - y!\left(\frac{27}{4}\right)^y.
\end{gather*}
We will show that $c(y) > 0 \text{ when } y > 0$. First, We note that $c(1) > 0$. We now use induction; assume that $c(y) > 0$ for some $y$. We will show that $c(y+1) > 0$. We have that 
\begin{gather*}
    c(y) = (2y+2)...(3y+2)-y!\left(\frac{27}{4}\right)^{y} > 0 \\
    \implies d(y) = \frac{(2y+2)...(3y+2)}{y!\left(\frac{27}{4}\right)^y} > 1.
\end{gather*}
We want to show that 
\begin{gather*}
    c(y+1) = (2y+4)...(3y+5) - (y+1)!\left(\frac{27}{4}\right)^{y+1} > 0 \\
    \Leftrightarrow d(y+1) = \frac{(2y+4)...(3y+5)}{(y+1)!\left(\frac{27}{4}\right)^{y+1}} > 1.
\end{gather*}
We will show that 
\begin{gather*}
    \frac{d(y+1)}{d(y)} = \frac{(3y+3)(3y+4)(3y+5)}{(2y+2)(2y+3)(y+1)\frac{27}{4}} > 1 \\ 
    \Leftrightarrow (3y+3)(3y+4)(3y+5) - (2y+2)(2y+3)(y+1)\frac{27}{4} > 0.
\end{gather*}
We note that the last expression turns out to be a quadratic in $y$ which is always positive when $y > 0$. Hence, $\frac{d(y+1)}{d(y)} > 1$ when $y > 0$. Thus, since by our induction hypothesis we have that $d(y) > 1, y > 0$ and by our proof above we have that $\frac{d(y+1)}{d(y)} > 1, y > 0$, we note that this implies $d(y+1) = d(y)\frac{d(y+1)}{d(y)} > 1$, which is what we wanted to show in our induction step. Hence, proved.

\subsubsection{Extending to General \texorpdfstring{$K$}{}}
For a general $K$, PPR-1v1 stops when 
\begin{gather*}
    \frac{\left(\frac{1}{2}\right)^{s_1^t+s_2^t}(s_1^t+s_2^t+1)!}{s_1^t!s_2^t!} \leq \frac{\delta}{K-1}
\end{gather*}
and when PPR-MD stops, it is true that (by choosing $x = \left(\frac{\hat{p}^t_1+\hat{p}^t_2}{2}, \frac{\hat{p}^t_1+\hat{p}^t_2}{2}, \hat{p}^t_3, ... \hat{p}^t_{K}\right)$
\begin{gather*}
    \frac{\left(\frac{\hat{p}^t_1+\hat{p}^t_2}{2}\right)^{s_1^t+s_2^t}(\hat{p}^t_3)^{s_3^t}(\hat{p}^t_4)^{s_4^t}...(\hat{p}^t_K)^{s_K^t}(s_1^t+s_2^t+s_3^t+s_4^t+...+s_K^t+K-1)!}{s_1^t!s_2^t!s_3^t!s_4^t!...s_K^t!} \leq \frac{\delta}{(K-1)!}.
\end{gather*}
If we show that 
\begin{gather*}
    \frac{(K-2)!\left(\frac{\hat{p}^t_1+\hat{p}^t_2}{2}\right)^{s_1^t+s_2^t}(\hat{p}^t_3)^{s_3^t}(\hat{p}^t_4)^{s_4^t}..(\hat{p}^t_K)^{s_K^t}(s_1^t+s_2^t+s_3^t+s_4^t+...+s_K^t+K-1)!}{s_1^t!s_2^t!s_3^t!s_4^t!...s_K^t!} \geq \\ \frac{\left(\frac{1}{2}\right)^{s_1^t+s_2^t}(s_1^t+s_2^t+1)!}{s_1^t!s_2^t!},
\end{gather*}    
then PPR-MD stopping will imply PPR-1v1 stopping. We will show this by using a sequence of inequalities as follows;
\begin{gather*}
    \frac{(K-2)!\left(\frac{\hat{p}^t_1+\hat{p}^t_2}{2}\right)^{s_1^t+s_2^t}(\hat{p}^t_3)^{s_3^t}(\hat{p}^t_4)^{s_4^t}..(\hat{p}^t_K)^{s_K^t}(s_1^t+s_2^t+s_3^t+s_4^t+...+s_K^t+K-1)!}{s_1^t!s_2^t!s_3^t!s_4^t!...s_K^t!} \geq \\ \frac{(K-3)!\left(\frac{\hat{p}^t_1+\hat{p}^t_2}{2}\right)^{s_1^t+s_2^t}(\hat{p}^t_3)^{s_3^t}...(\hat{p}^t_{K-1})^{s_{K-1}^t}(s_1^t+s_2^t+s_3^t+...+s_{K-1}^t+K-2)!}{s_1^t!s_2^t!s_3^t!...s_{K-1}^t!} \geq ... \geq \\ \frac{\left(\frac{1}{2}\right)^{s_1^t+s_2^t}(s_1^t+s_2^t+1)!}{s_1^t!s_2^t!}.
\end{gather*}   
We will show the first inequality, and the rest follow the same structure except the last (for $K = 3$), which was already shown in the subsection above. Note that the last inequality does not follow the same structure due to the lack of $\frac{\hat{p}^t_1+\hat{p}^t_2}{2}$ term in the RHS, since this does not show up in the PPR-1v1 expression. Showing the first inequality is equivalent to showing
\begin{gather*}
    (K-2)(\hat{p}^t_K)^{s_K^t} \geq \frac{(s_1^t+s_2^t+...+s_{K-1}^t+K-2)!s_K^t!}{(s_1^t+s_2^t+...+s_{K}^t+K-1)!} \\
    \Leftrightarrow (K-2)\frac{(s_K^t)^{s_K^t}}{(s_1^t+s_2^t+...+s_K^t)^{s_K^t}} \geq \frac{(s_1^t+s_2^t+...+s_{K-1}^t+K-2)!s_K^t!}{(s_1^t+s_2^t+...+s_{K}^t+K-1)!} \\
    \Leftrightarrow (K-2)\frac{y^{y}}{(x+y)^{y}} \geq \frac{(x+K-2)!y!}{(x+y+K-1)!}, x \geq (K-1)y, \text{ where } x = s_1^t+s_2^t+..+s_{K-1}^t, y = s_K^t. 
\end{gather*}
We already know that \begin{gather*}
    \frac{x^{x}y^{y}}{(x+y)^{x+y}} \geq \frac{(x+1)!y!}{(x+y+2)!}, \text{ where } x \geq 2y, y > 0
\end{gather*}
from our proof for $K = 3$. In this equation, mutltiplying the LHS by $\frac{(K-2)(x+y)^{x}}{x^{x}}$ which is a factor greater than $1$, and multiplying the RHS by $\frac{(x+2)...(x+K-2)}{(x+y+3)...(x+y+K-1)}$, which is a factor lesser than $1$, we have the inequality that 
\begin{gather*}
    \frac{x^{x}}{(x+y)^{y}} \geq \frac{(x+K-2)!y!}{(x+y+K-1)!}, x \geq 2y
\end{gather*}
which directly implies what we wanted to show. Hence, even in the case of $K$ parties, PPR-MD stopping implies PPR-1v1 stopping.

\subsection{PPR-1v1 and \texorpdfstring{$\mathcal{A}_1$}{}-1v1}
\label{app:ppr-a1}

Both PPR-1v1 and $\mathcal{A}_1$-1v1 reduce to a single application of PPR-Bernoulli and $\mathcal{A}_1$, respectively, between $v_{\text{first}(t)}$ and $v_{\text{second}(t)}$. Thus it suffices to prove $\mathcal{G}(\text{PPR-Bernoulli}, \mathcal{A}_{1}, X, \delta)$, assuming $X$ is generated by distribution $\mathcal{P}$ over two values.

After any arbitrary $t$ timesteps, without loss of generality, assume $\text{first}(t) = 1$ and $\hat{p}_1^t = \frac{s_1^t}{t}$. We prove the following:
\begin{align*}
\neg TERM_{PPR}^t &\implies \neg TERM_{\mathcal{A}1}^t, \text{ where}\\
TERM_{PPR}^t &\iff \text{Beta}(0.5;s_1^t+1,t-s_1^t+1) \leq \delta, \text{ and}\\
TERM_{\mathcal{A}1}^t &\iff \hat{p}_1^t - \beta(t, \delta) > 0.5, \qquad \beta(t, \delta) = \sqrt{\frac{2V^t\text{ln}(4 t^2/\delta)}{t}} + \frac{7\text{ln}(4 t^2/\delta)}{3(t-1)}. \nonumber
\end{align*}


From Lemma~\ref{lem:term} (see Appendix~\ref{app:optimality-final}), $\neg TERM_{PPR}^t$ implies,
\begin{equation*}
    tKL\left(\hat{p}_1^t||0.5\right) < \ln\left(\frac{t+1}{\delta}\right).
\end{equation*}
Applying Pinsker's inequality, we get,
\begin{equation}
    \label{eqn:pinsker}
    t\frac{(2(\hat{p}_1^t - 0.5))^2}{2\ln(2)} < \ln\left(\frac{t+1}{\delta}\right).\\  
\end{equation}
\paragraph{Case 1:} $0.50 \leq \hat{p}_1^t \leq 0.77$ \\
Rearranging equation \ref{eqn:pinsker} gives us,
\begin{align*}
    & \hat{p}_1^t < 0.5 + \sqrt{\frac{\ln(2)}{2t}\ln\left(\frac{t+1}{\delta}\right)} \\
    \implies & \hat{p}_1^t < 0.5 + \sqrt{\frac{2\hat{p}_1^t(1-\hat{p}_1^t)\text{ln}(4 t^2/\delta)}{(t-1)}}  \\
    \implies & \hat{p}_1^t < 0.5 + \left(\sqrt{\frac{2\hat{p}_1^t(1-\hat{p}_1^t)\text{ln}(4 t^2/\delta)}{(t-1)}} + \frac{7\text{ln}(4 t^2/\delta)}{3(t-1)} \right) \\
    \implies & \hat{p}_1^t < 0.5 + \beta(t,\delta) \\
    \implies & \neg TERM_{\mathcal{A}1}^t.
\end{align*}
\paragraph{Case 2:} $0.77 < \hat{p}_1^t \leq 1$ \\
The lower confidence bound given by the $\mathcal{A}_1$ algorithm is
\begin{align*}
    \hat{p}_1^t - \beta(t, \delta) &= \hat{p}_1^t - \left(\sqrt{\frac{2V^t\text{ln}(4 t^2/\delta)}{t}} +
    \frac{7\text{ln}(4 t^2/\delta)}{3(t-1)}\right) \\
    \implies \hat{p}_1^t - \beta(t, \delta) &\leq \hat{p}_1^t - \frac{7}{3}\ln\left(\frac{t+1}{\delta}\right) \\
    \implies \hat{p}_1^t - \beta(t, \delta) &\leq \hat{p}_1^t - \frac{14(0.5-\hat{p}_1^t)^2}{3\ln(2)}\quad \text{(using equation \ref{eqn:pinsker}).}\\
\end{align*}
Since $f(\hat{p}_1^t) = \hat{p}_1^t - \frac{14(0.5-\hat{p}_1^t)^2}{3\ln{2}}$ is a decreasing function for $\hat{p}_1^t \geq 0.77$, and $f(0.77) = 0.28$, we have $\neg TERM_{PPR}^t$ implies
\begin{align*}
     &\hat{p}_1^t - \beta(t, \delta) \leq 0.5 \\
     \implies &\neg TERM_{\mathcal{A}1}^t.
 \end{align*}

%% file: appendices/optimality_final.tex
\newpage

\section{PROOF OF ASYMPTOTIC OPTIMALITY OF PPR-1v1}
\label{app:optimality-final}

We present a full proof of Theorem~\ref{thm:asymptotic}. Consider a categorical distribution $\mathcal{P}(p,v,K)$ having parameters $p_1 > p_2 \ldots \geq p_K$ for $K \geq 2$. We show that PPR-1v1 asymptotically matches the lower bound given in Theorem \ref{thm:lowerbound} when $\mathcal{P}'$ is defined as follows: $p_1' = p_2' = \frac{p_1+p_2}{2}$ and $\forall i>2, p_i' = p_i$. Observe that although $\mathcal{P}'$ itself does not have $v_{2}$ as its mode, it yields the supremum defined in $\text{LB}(\mathcal{P}, \delta)$ among distributions that do have $v_{2}$ as mode. Hence it is a sufficient choice for our proof,
which proceeds in two major steps.

\begin{enumerate}
    \item In subsection \ref{app:D2} we fix arbitrary $\epsilon > 0$ as input, and define $T^*$ such that $\lim\limits_{\delta\rightarrow0}\frac{T^*}{\ln(1/\delta)} = \frac{(1+\epsilon)}{KL(\mathcal{P}||\mathcal{P}')}$. Then assuming we have $T^*$ samples in total, we lower bound the number of samples obtained for each of the 1v1 tests involving $v_1$.
    \item In subsection \ref{App:D2}, we show that the number of samples obtained for each test is sufficient for PPR-Bernoulli to terminate and declare $v_1$ as the winner. The arguments in this subsection are along the lines of those used by \cite{pmlr-v49-garivier16a}.
\end{enumerate}

\subsection{Lower Bounding the Number of Samples Obtained for each 1v1 Test}
\label{app:D2}

Given arbitrary $\epsilon > 0$ and $\alpha > 0$, let us introduce $T^*$:

$$T^* \eqdef \frac{1+\epsilon}{KL(\mathcal{P}||\mathcal{P}')}\left(\ln\left(\frac{K-1}{\delta}\right) + \max\limits_{j\in \{2, 3\ldots K\}} H_j\right) + \frac{K}{1-K\xi} N(\mathcal{P},\xi,\epsilon,\alpha) + \frac{K}{1-K\xi}T_0(\gamma).$$

We start by defining the notation used:
\begin{itemize}
    \item $\xi(\mathcal{P}, \epsilon)$ : $0 < \xi < \min|p_1-p_2|/4$ is small enough so that $\left(1 - \frac{\xi}{p_1 + p_K}\right)(1 + \frac{\epsilon}{2}) > 1$.
    \item $\gamma(\epsilon)$: $\gamma > 0$ is small enough so that $(1+\gamma)(1+\frac{\epsilon}{2}) < (1 + \epsilon)$.
    \item $T_0(\gamma)$ : $T_0$ is large enough such that $\forall t \geq T_0(\gamma), t^{1+\gamma} > t + 1$.
    \item $H_j = \ln\left(e\left(\frac{(1+\epsilon/2)(1 - \xi/(p_1 + p_j)}{KL\left(p_1/(p_1 + p_j) || 0.5\right)}\right)^{1+\gamma} \right) + \ln\left(\ln\left(\frac{K-1}{\delta}\left(\frac{(1+\epsilon/2)(1 - \xi/(p_1 + p_j)}{KL\left(p_1/(p_1 + p_j) || 0.5\right)}\right)^{1+\gamma} \right)\right)$.
    \item  $\xi'(\mathcal{P}, \xi, \epsilon)$: $\xi'$ is such that  $|\hat{p}_i^t - p_i| < \xi' \implies D\left(\frac{\hat{p}_1^t}{\hat{p}_1^t + \hat{p}_i^t} || 0.5\right) \geq \frac{D(p_1/(p_1+p_i) || 0.5)}{\left(1 - \xi/(p_1 + p_K)\right)(1 + \epsilon/2)}$ for $i \in \{2,3,\ldots,K\}$
    \item $N(\mathcal{P}, \xi,  \epsilon, \alpha)$ : First, we introduce $\sigma = \max\left\{t \in N^*: |\hat{p}_i^t - p_i| \geq \min(\frac{\xi}{2}, \xi') \right\}$. By the law of large numbers, for every $\alpha \in (0,1)$, there exists $N(\mathcal{P}, \xi,  \epsilon, \alpha)$, such that $\mathbb{P} \left(\sigma \leq N(\mathcal{P}, \xi,  \epsilon, \alpha)\right) > 1-\alpha$.
    \item $E_\alpha$: We define the event $E_\alpha = \left\{\forall t \geq N(\mathcal{P}, \xi,  \epsilon, \alpha), \forall i \in \{1,2 \ldots K\} \quad |\hat{p}_i^t - p_i| < \min(\frac{\xi}{2}, \xi') = \xi''\right\}$. From the definition of $N$, $\mathbb{P}(E_\alpha) > 1 - \alpha$.
\end{itemize}

At time step $t > 0$ let $s_i^t$ denote the number of samples of $v_i$, and $\hat{p}_i^t = \frac{s_i^t}{t}$ denote the empirical mean for value $v_{i}$. For convenience we use $KL(q_1||q_2)$ to denote $KL(\text{Bernoulli}(q_1)||\text{Bernoulli}(q_2))$. We will now show that conditioned on $E_\alpha$, we have enough samples to separate each class $i \in \{2, 3, \dots, K\}$ from the mode 1, i.e. $s_1^t + s_i^t$ is sufficiently large.

\begin{align*}
    s_1^{T^*} + s_i^{T^*} &= (\hat{p}_1^{T^*} + \hat{p}_i^{T^*})T^* \\
    &\geq (p_1 + p_i -\xi)T^* \\
    &= \frac{(p_1 + p_i)\left(1 - \xi/(p_1+p_i)\right)(1 + \epsilon)}{KL(\mathcal{P}||\mathcal{P}')}\left(\ln\left(\frac{K-1}{\delta}\right) + \max\limits_{j\in \{2,3\ldots K\}}H_j\right) \\ 
    & + K(p_1 + p_i)\left(\frac{1 - \xi/(p_1+p_i)}{1-K\xi}\right)(N(\mathcal{P}, \xi, \epsilon, \alpha) + T_0(\gamma))\\
    &\geq \frac{\left(1 - \xi/(p_1+p_i)\right)(1 + \epsilon)}{KL(p_1/(p_1 + p_i)||0.5)}\left(\ln\left(\frac{K-1}{\delta}\right) + H_i\right) + N(\mathcal{P}, \xi, \epsilon, \alpha) + T_0(\gamma)\\ 
    &\geq \frac{\left(1 - \xi/(p_1+p_i)\right)(1 + \epsilon/2)(1+\gamma)}{KL(p_1/(p_1 + p_i)||0.5)}\left(\ln\left(\frac{K-1}{\delta}\right) + H_i\right) + N(\mathcal{P}, \xi, \epsilon, \alpha) + T_0(\gamma)
\end{align*}

where the second inequality follows from Lemma \ref{lem:kl}.

\subsection{Termination of the 1v1 Tests}
\label{App:D2}

We have shown that conditional on the event $E_\alpha$, 

$$s_1^{T^*} + s_i^{T^*} \geq \frac{\left(1 - \xi/(p_1+p_i)\right)(1 + \epsilon/2)(1+\gamma)}{KL(p_1/(p_1 + p_i)||0.5)}\left(\ln\left(\frac{K-1}{\delta}\right) + H_i\right) + N(\mathcal{P}, \xi, \epsilon, \alpha) + T_0(\gamma).$$
Consider $c_1 = \frac{KL(p_1/(p_1+p_i)||0.5)}{(1-\xi/(p_1+p_i))(1+\epsilon/2)}, c_2 = \frac{K-1}{\delta}, \alpha = 1 + \gamma$. Since for $p\in(0.5,1], KL(p||0.5) \leq \ln(2) < 1$ we have $c_1^\alpha \leq \ln(2)$. If $\delta \leq 0.5$, we can see that $\frac{c_2}{c_1^\alpha} \geq \frac{2}{\ln(2)} > e$. This allows us to use Lemma \ref{lem:technical} and state the following for $\delta \leq 0.5$,

\begin{align*}
(s_1^{T^*} + s_i^{T^*})\frac{KL(p_1/(p_1+p_i)||0.5)}{(1-\xi/(p_1+p_i))(1+\epsilon/2)} &\geq \ln\left(\frac{(K-1)(s_1^{T^*} + s_i^{T^*})^{1+\gamma}}{\delta}\right) \\
& \geq \ln\left(\frac{(K-1)(s_1^{T^*} + s_i^{T^*}+1)}{\delta}\right).
\end{align*}

By the definition of $N(\mathcal{P}, \xi,  \epsilon, \alpha)$, for $\delta \leq 0.5$ we know that,

\begin{align*}
(s_1^{T^*_i} + s_i^{T^*})KL(\hat{p}^{T^*}_1/(\hat{p}^{T^*}_1+\hat{p}^{T^*}_i)||0.5) &\geq \ln\left(\frac{(K-1)(s_1^{T^*} + s_i^{T^*}+1)}{\delta}\right) \\
\implies \text{Beta}(0.5;s_1^{T^*}+1, s_i^{T^*} + 1) &\leq \frac{\delta}{K-1} \\
\end{align*}

where the last implication follows from Lemma \ref{lem:term}.
At this stage the PPR-Bernoulli Stopping Rule declares that $p_1^{T^*} > p_i^{T^*}$.

The above proof tells us that under the event $E_\alpha$, after $T^*$ samples, PPR-Bernoulli decides that for $i \in \{2, 3, \dots, K\}$, $p_1 > p_i$, i.e. $v_1$ is the mode of the distribution. Therefore, if $\tau$ denotes the stopping time of PPR-1v1, we have the following bound.

\begin{align*}
\mathbb{P}(\tau \leq T^*) > 1 - \alpha
\implies
\mathbb{P}
\left(\lim_{\delta\rightarrow0}\frac{\tau}{\ln(1/\delta)} \leq \frac{1+\epsilon}{KL\left(\mathcal{P} || \mathcal{P'}\right)} \right) &> 1 - \alpha.
\end{align*}


To show that the lower bound of \cite{Shah_Choudhury_Karamchandani_Gopalan_2020} is matched, we upper bound the expected stopping time, using a commonly-known\footnote{See Theorem 2.1 by \cite{mulzer2019proofs}.} result to bound the tail probability for a binomial random variable.
\begin{align*}
    \mathbb{P}(\tau > T^*) &\leq \mathbb{P}(E^\complement_\alpha). \\
    \mathbb{E(\tau)} &\leq T^* + \sum\limits_{i=1}^{K}\sum\limits_{t=N(\mathcal{P}, \xi, \epsilon, \alpha)}^\infty \mathcal{P}(|\hat{p}_i^t - p_i| > \xi'') \\
    &\leq T^* + \sum\limits_{i=1}^{K}\sum\limits_{t=N(\mathcal{P}, \xi, \epsilon, \alpha)}^\infty \exp\left(-2tKL(p_i-\xi''||\xi)\right)  \\
    &\leq T^* + \sum\limits_{i=1}^{K}\sum\limits_{t=0}^\infty \exp\left(-2tKL(p_i-\xi''||\xi)\right)  \\
    &\leq T^* + \sum\limits_{i=1}^{K}\frac{1}{1 - \exp(2KL(p_i - \xi''||p_i)}.
\end{align*}

Hence, we have $$\lim_{\delta\rightarrow0}\frac{\mathbb{E}_{\mathcal{P}}(\tau)}{\ln(1/\delta)} \leq \frac{1+\epsilon}{KL\left(\mathcal{P} || \mathcal{P'}\right)},$$ which completes our proof.

We now furnish proofs of the lemmas that were used in the proofs above.

\begin{lemma}
\label{lem:kl}
For $x \in [0,p_2]$, $$(p_1+x)KL\left(\frac{p_1}{p_1+x} || 0.5\right) \geq KL(\mathcal{P}||\mathcal{P}').$$
\end{lemma}

\begin{proof}
By definition,

\begin{align*}
(p_1+x)KL\left(\frac{p_1}{p_1+x} || 0.5\right) &= p_1\ln\left(\frac{p_1}{(p_1+x)/2}\right) + x\ln\left(\frac{x}{(p_1+x)/2}\right) \\
&\geq p_1\ln\left(\frac{p_1}{(p_1+p_2)/2}\right) + p_2\ln\left(\frac{p_2}{(p_1+p_2)/2}\right) \\
&= KL(\mathcal{P}||\mathcal{P}').
\end{align*}

The second line follows from from the fact that $p_2 \geq x$ and that $F(x) = p_1\ln\left(\frac{p_1}{(p_1+x)/2}\right) + p\ln\left(\frac{x}{(p_1+x)/2}\right), x \in [0, p_2]$ is a decreasing function ($F'(x) = \ln\left(\frac{2x}{p_1 + x}\right) < 0$).
\end{proof}

\begin{lemma}
\label{lem:technical}
For $\alpha \in [1,e/2]$ for any two constants $c_1$ and $c_2$ such that $\frac{c_2}{c_1^\alpha} \geq e$, we have 
$$c_1x \geq \ln(c_2x^\alpha), \forall x \geq T_0(c_1, c_2, \alpha) \text{ where } T_0(c_1, c_2, \alpha) = \frac{\alpha}{c_1}\left[\ln\left(\frac{c_2e}{c_1^\alpha}\right) + \ln\ln\left(\frac{c_2}{c_1^\alpha}\right)\right].$$
\end{lemma}
\begin{proof}
Consider $F(x) = c_1x - \ln\left(c_2x^\alpha\right)$. We have $F'(x) = c_1 - \frac{\alpha}{x}$ which means that $F(x)$ is increasing for $x > \frac{\alpha}{c_1}$.

Using Lemma 18 by \cite{pmlr-v49-garivier16a} we know that for $x = T_0(c_1, c_2, \alpha), c_1x \geq c_2x^\alpha$, and $\frac{c_2}{c_1^\alpha} \geq e \implies T_0(c_1, c_2, \alpha) \geq \frac{\alpha}{c_1}[\ln(e^2) + \ln\ln(e)] = 2\frac{\alpha}{c_1}$. The fact that $F(x)$ is increasing at this value of $x$ completes our proof.
\end{proof}

\begin{lemma}
\label{lem:term}
For $i \in \{2,3,\ldots,K\}$,
$$(s_1^{t} + s_i^{t})KL(\hat{p}^{t}_1/(\hat{p}^{t}_1+\hat{p}^{t}_i)||0.5) \geq \ln\left(\frac{s_1^t + s_i^t +1}{\delta}\right) 
\implies \text{Beta}(0.5;s_1^{t}+1, s_i^{t} + 1) \leq \delta.$$
\end{lemma}

\begin{proof}

\begin{align*}
    (s_1^{t} + s_i^{t})KL(\hat{p}^{t}_1/(\hat{p}^{t}_1+\hat{p}^{t}_i)||0.5) &\geq \ln\left(\frac{s_1^t + s_i^t +1}{\delta}\right) \\
    \implies (s_1^t + s_i^t)\frac{\hat{p}_1^t}{\hat{p}_1^t + \hat{p}_i^t}\ln\left(\frac{\hat{p}_1^t}{\hat{p}_1^t + \hat{p}_i^t}\right) + (s_1^t + s_i^t)\frac{\hat{p}_i^t}{\hat{p}_1^t + \hat{p}_i^t}\ln\left(\frac{\hat{p}_i^t}{\hat{p}_1^t + \hat{p}_i^t}\right)  - (s_1^t+s_i^t)\ln(0.5) &\geq \ln\left(\frac{s_1^t + s_i^t +1}{\delta}\right) \\
    \implies (s_1^t + s_i^t) H_b\left(\frac{\hat{p}_i^t}{\hat{p}_1^t+\hat{p}_i^t}\right) + (s_1^t+s_i^t)\log_2(0.5) &\leq \log_2\left(\frac{\delta}{s_1^t + s_i^t +1}\right). 
\end{align*}

The last step follows from the definition of binary entropy function $H_b$ for $p \in [0, 1]$: 
\begin{equation*}
    H_b(p) = 
    \begin{cases}
    0 & p \in \{0, 1\},\\
    -p\log_2p - (1 - p)\log_2(1 - p) & \text{otherwise.}\\
    \end{cases}
\end{equation*}

Writing the condition in this form allows us to use the following well known\footnote{See, for example, \cite{galvin2014}.} inequality, with $\alpha = \frac{\hat{p}_i^t}{\hat{p}_1^t+\hat{p}_i^t}$  ($\alpha < \frac{1}{2}$ as $s_1^t > s_i^t$) and $t=s_1^t + s_i^t$.

\begin{equation*}
    \binom{t}{\alpha t} \leq 2^{tH_b(\alpha)}\,\, \text{where $t \in \mathbb{N}$ and $\alpha \in \left[0, \frac{1}{2}\right].$} 
\end{equation*}

So we have

\begin{align*}
    (s_1^t + s_i^t) H_b\left(\frac{\hat{p}_1^t}{\hat{p}_1^t+\hat{p}_i^t}\right) + (s_1^t+s_i^t)\log_2(0.5) &\leq \log_2\left(\frac{\delta}{s_1^t + s_i^t +1}\right) \\
    \implies \log_2\left(\binom{s_1^t + s_i^t}{s_1^t}\right) + (s_1^t+s_i^t)\log_2(0.5) + \log_2(s_1^t + s_i^t +1) &\leq \log_2\left(\delta\right) \\
    \implies \log_2\left(\frac{(s_1^t + s_i^t + 1)!(0.5)^{s_1^t}(0.5)^{s_i^t}}{(s_1^t)!(s_i^t)!}\right) &\leq \log_2\left(\delta\right) \\
    \implies \text{Beta}(0.5; s_1^t + 1, s_i^t + 1) &\leq \delta.
\end{align*}

\end{proof}

%% file: appendices/proofoflemma3.tex
\newpage

\section{NON-ASYMPTOTIC UPPER BOUND FOR PPR-1v1}
\label{app:ppr-bernoulli-upperboundproof}

In Appendix~\ref{app:optimality-final} we showed that the PPR-1v1 stopping rule is asymptotically optimal, in the regime that $\delta \to 0$. In this section, we show a non-asymptotic upper bound on its sample complexity: in other words, a result that holds for all $\delta \in (0, 1)$. To the best of our knowledge, the tightest such upper bound given yet for the PAC mode estimation problem is the recent result of \citet{Shah_Choudhury_Karamchandani_Gopalan_2020}, which we reproduce below.

\begin{theorem}[$\mathcal{A}_{1}$ upper bound~\cite{Shah_Choudhury_Karamchandani_Gopalan_2020}]
\label{thm:a1upperbound}
Fix $\delta \in (0, 1)$, $K \geq 2$, and problem instance $\mathcal{P} = (p, v, K)$. When $\mathcal{A}_{1}$ is run on $\mathcal{P}$, with probability $1 - \delta$, the number of samples it observes is at most $$\frac{592}{3}\frac{p_1}{(p_1-p_2)^2}\ln\left(\frac{592}{3}\sqrt{\frac{K}{\delta}}\frac{p_1}{(p_1-p_2)^2}\right).$$
\end{theorem}

It is easy to show that the leading $\frac{p_{1}}{(p_{1} - p_{2})^{2}}$ factor is within a constant factor of $LB(\mathcal{P}, \delta)$~\citep{Shah_Choudhury_Karamchandani_Gopalan_2020}. In this appendix, we derive a similar upper bound for PPR-1v1, albeit one that is tighter by a small constant factor. Our first step is to show an upper bound for the special case of $K = 2$ (wherein PPR-1v1 reduces to PPR-Bernoulli) in Appendix~\ref{app:ubk2}. In turn, this result is used to generalise to $K \geq 2$ in Appendix~\ref{app:ubgenk}. The final upper bound, is given in Theorem~\ref{thm:ppr-meupperbound}.

\subsection{An Upper Bound for K = 2}
\label{app:ubk2}

\begin{lemma}[PPR-Bernoulli upper bound]

\label{lem:ppr-bernoulli-upperbound}
Fix $\delta \in (0, 1)$ and problem instance $\mathcal{P} = (p, v, 2)$. When PPR-Bernoulli is run on $\mathcal{P}$, with probability $1 - \delta$, the number of samples it observes is at most $$\frac{20.775 p_1}{(p_1-\frac{1}{2})^2}\ln\left(\frac{2.49}{(p_1-\frac{1}{2})^2 \delta}\right).$$
\end{lemma}

\begin{proof}
For a problem instance $\mathcal{P} = (p, v, 2)$, our parameters are $p_1$ and $p_2 = 1-p_1$, with $p_1 > 0.5$. It suffices to maintain a confidence sequence on $p_{1}$; termination is achieved when this confidence sequence no longer contains $\frac{1}{2}$. To upper-bound the number of samples needed for termination, we proceed in three steps.


\begin{enumerate}
\item At time $t$, there are $2^t$ possible 0-1 sequences that the Bernoulli variable can produce in $t$ steps. Let $X^{t}$ be a random variable denoting this $t$-length 0-1 sequence. Let $s_1^t$ denote the number of times $v_1$ occurs in $X^{t}$. In
Subsection~\ref{appA:1}, we find the range of $s_1^t$ for which  $\mathds{1}\left\{R^t(\frac{1}{2}) \geq \frac{1}{\delta}\right\}$.
\item Next, in Appendix~\ref{appA:2}, we use the range of $k$ derived in Appendix~\ref{appA:1} to derive a sufficient condition for $t$ to be the sample complexity.
\item In Appendix~\ref{appA:3}, we use the sufficient condition derived in Subsection \ref{appA:2} to obtain a closed-form sample complexity upper bound. We separately take up two cases, $p_1 \leq 0.6$, and $p_1 > 0.6$, so as to tighten the constants in the upper bound.
\end{enumerate}

\subsubsection{Finding a Range of \texorpdfstring{$k$}{} for which  \texorpdfstring{$\mathds{1}\left\{R^t(\frac{1}{2}) \geq \frac{1}{\delta}\right\}$}{}}
\label{appA:1}
 The stopping rule does not terminate at $t$ so long as $R^t\left(\frac{1}{2}\right) < \frac{1}{\delta}$.
\begin{gather*}
\mathbb{P}\left(R^t\left(\frac{1}{2}\right) \geq \frac{1}{\delta}\right) = \sum_{X^t}\mathds{1}\left\{R^t\left(\frac{1}{2}\right) \geq \frac{1}{\delta}\right\} P_{p_1}(X^t).
\end{gather*}
The expression inside the indicator random variable can be simplified to obtain the following:
\begin{align*}
\mathds{1}\left\{R^t\left(\frac{1}{2}\right) \geq \frac{1}{\delta}\right\} = \mathds{1}\left\{\frac{1}{\pi_t(1/2)} \geq \frac{1}{\delta}\right\} = \mathds{1}\left\{\frac{\int_{\eta = 0}^{1} P_{\eta}(X^t)d\eta}{P_{1/2}(X^t)} \geq \frac{1}{\delta}\right\}.
\end{align*}

The sum inside the indicator function is independent of the position of 0-1 outcomes within the sequence $X^t$ and only depends on the number of zeros and ones in the instance. Suppose we consider the instances which contain $s_1^t$ ones; then, we have
\begin{align*}
\mathds{1}\left\{R^t\left(\frac{1}{2}\right) \geq \frac{1}{\delta}\right\} = \mathds{1}\left\{\delta \geq \frac{(t + 1)!}{s_1^t!(t-s_1^t)!}\left(\frac{1}{2}\right)^{s_1^t} \left(\frac{1}{2}\right)^{t - s_1^t}\right\}.
\end{align*}
From the above equations, we derive the exact expression for $\mathbb{P}(R^t(\frac{1}{2}) \geq \frac{1}{\delta})$:
\begin{gather*}
    \mathbb{P}\left(R^t\left(\frac{1}{2}\right) \geq \frac{1}{\delta}\right) = \sum_{s_1^t=0}^t \binom{t}{s_1^t} p_1^{s_1^t} (1 - p_1)^{t-s_1^t}\mathds{1}\left\{\delta \geq \frac{(t + 1)!}{s_1^t!(t-s_1^t)!}2^{-t}\right\}.
\end{gather*}
We now break the expression into two parts.
\begin{align*}
    \mathbb{P}\left(R^t\left(\frac{1}{2}\right) \geq \frac{1}{\delta}\right) &= \sum_{s_1^t=0}^t \binom{t}{s_1^t} p_1^{s_1^t} (1 - p_1)^{t-s_1^t}\mathds{1}\left\{\delta \geq \frac{(t + 1)!}{s_1^t!(t-s_1^t)!}2^{-t}\right\}\\
    = &\sum_{s_1^t=0}^{\left \lfloor{\frac{t}{2}}\right \rfloor} \binom{t}{s_1^t} p_1^{s_1^t} (1 - p_1)^{t-s_1^t}\mathds{1}\left\{\frac{\delta}{t + 1} \geq \binom{t}{s_1^t}2^{-t}\right\} + \\
    &\sum_{s_1^t=\left \lfloor{\frac{t}{2}}\right \rfloor + 1}^{t} \binom{t}{s_1^t} p_1^{s_1^t} (1 - p_1)^{t-s_1^t}\mathds{1}\left\{\frac{\delta}{t + 1} \geq \binom{t}{t - s_1^t}2^{-t}\right\}.
\end{align*}
We have split the summation in this manner as it allows to use the following commonly-known\footnote{See, for example, \cite{galvin2014}.} inequality for the binomial coefficients inside the indicator function.
\begin{equation*}
    \binom{t}{\alpha t} \leq 2^{tH_b(\alpha)}\,\, \text{where $t \in \mathbb{N}$ and $\alpha \in \left[0, \frac{1}{2}\right].$} 
\end{equation*}
The definition of binary entropy function $H_b$ for $p \in [0, 1]$ is: 
\begin{equation*}
    H_b(p) = 
    \begin{cases}
    0 & p \in \{0, 1\},\\
    -p\log_2p - (1 - p)\log_2(1 - p) & \text{otherwise.}\\
    \end{cases}
\end{equation*}
Invoking this inequality in the probability expression, we arrive at the lower bound:
\begin{align*}
    \mathbb{P}\left(R^t\left(\frac{1}{2}\right) \geq \frac{1}{\delta}\right) \geq  &\sum_{s_1^t=0}^{\left \lfloor{\frac{t}{2}}\right \rfloor} \binom{t}{s_1^t} p_1^{s_1^t} (1 - p_1)^{t-s_1^t}\mathds{1}\left\{\frac{1}{t}\log_2\left(\frac{\delta}{t + 1}\right) + 1 \geq H_b\left(\frac{s_1^t}{t}\right)\right\} + \\
    &\sum_{s_1^t=\left \lfloor{\frac{t}{2}}\right \rfloor + 1}^{t} \binom{t}{s_1^t} p_1^{s_1^t} (1 - p_1)^{t-s_1^t}\mathds{1}\left\{\frac{1}{t}\log_2\left(\frac{\delta}{t + 1}\right) + 1 \geq H_b\left(\frac{t - s_1^t}{t}\right)\right\}.
\end{align*}
Using the fact that
\[H_b(p) \leq 2\sqrt{p(1 - p)} \,\,\text{if}\,\, p \leq \frac{1}{2}, \]
the summation is further lower-bounded:
\begin{align*}
    \mathbb{P}\left(R^t\left(\frac{1}{2}\right) \geq \frac{1}{\delta}\right) \geq  &\sum_{s_1^t=0}^{t} \binom{t}{s_1^t} p_1^{s_1^t} (1 - p_1)^{t-s_1^t}\mathds{1}\left\{\frac{1}{2}\left[\log_2\left(\frac{\delta}{t + 1}\right) + t\right] \geq \sqrt{s_1^t(t - s_1^t)}\right\}.
\end{align*}
We will later see that for the $t$ we find as the bound, $\log_2\left(\frac{\delta}{t + 1}\right) + t \geq 0$. Hence, we can take the square of both sides. By using that fact and solving the quadratic, the range of $s_1^t$ which satisfies the condition inside the indicator function is:
\[s_1^t \in \left[0, \frac{t - \sqrt{t^2 - 4\beta^2}}{2}\right] \bigcup \left[\frac{t + \sqrt{t^2 - 4\beta^2}}{2}, t\right] \,\, \text{where} \,\, \beta = \frac{1}{2}\left[\log_2\left(\frac{\delta}{t + 1}\right) + t\right].\] 
Slightly loosening this bound, we get
\begin{align*}
    s_1^t \in \left[0, \frac{t - \sqrt{2t\log_2\left(\frac{t+1}{\delta}\right)}}{2}\right] \bigcup \left[\frac{t + \sqrt{2t\log_2\left(\frac{t+1}{\delta}\right)}}{2}, t\right].
\end{align*}
In short, if $s_1^t$ lies in the range above, then it must satisfy $\mathds{1}\left\{R^t(\frac{1}{2}) \geq \frac{1}{\delta}\right\}$. Call this range $R$.


\subsubsection{Finding a Sufficient Condition for \texorpdfstring{$t$}{} to be the Sample Complexity}
\label{appA:2}
We will lower bound the probability of $\sum_{s_1^t \in R}p_1^{s_1^t}(1-p_1)^{t-s_1^t}$ by $1-\delta$, and thus find the values of $t$ satisfying this equation. 
To that end, we use another commonly-known~\footnote{See Theorem 2.1 in \cite{mulzer2019proofs}.} result to bound the tail probability for a binomial random variable: for $\epsilon > 0$,
\begin{gather*}
    \mathbb{P}(H(t) \leq (p_1-\epsilon)t) \leq e^{-D_{KL} (p_1-\epsilon || p_1)t }.
\end{gather*}
Using this in our working,  considering $H(t)$ as a binomial random variable, we get
\begin{gather*}
    \sum_{s_1^t=(p_1-\epsilon)t}^{t} \binom{t}{s_1^t} p_1^{s_1^t} (1 - p_1)^{t-s_1^t} = \mathbb{P}\left(\left(p_1-\epsilon\right)t \leq H(t)  \right) \\ 
    \geq  1 - e^{-D_{KL} (p_1-\epsilon || p_1)t }  \\
    \implies \mathbb{P}\left(R^t\left(\frac{1}{2}\right) \leq \frac{1}{\delta} \right) \leq e^{-D_{KL} (p_1-\epsilon || p_1)t }.
\end{gather*}
We want $\mathbb{P}\left(R^t\left(\frac{1}{2}\right) > \frac{1}{\delta}\right) \geq 1 - \delta$. A sufficient condition for the above to hold is
\begin{gather*}
    e^{-D_{KL} (p_1-\epsilon || p_1)t } \leq \delta \\
    \implies t \times  D_{KL} (p_1-\epsilon || p_1) \geq \ln\left(\frac{1}{\delta}\right) 
\end{gather*}

And $\epsilon$ is given by, 

\begin{gather*}
    \epsilon = p_1 - \frac{1}{2}\left(1 + \sqrt{\frac{2\log_2\left(\frac{t+1}{\delta}\right)}{t}} \right), \epsilon \geq 0.
\end{gather*}
We will note that $\epsilon \geq 0$ holds later. For now, we try to find stronger sufficient conditions on $t$. To that end, we use that 
\begin{gather*}
    D_{KL}(x || y) \geq \frac{(x-y)^2}{2y}, x \leq y\\
    \implies D_{KL}\left(p_1-\epsilon || p_1\right) \geq \frac{\epsilon^2}{2p_1}.
\end{gather*}

So, a sufficient condition for 
\begin{gather*}
    t \times D_{KL}(p_1-\epsilon || p_1) \geq \ln\left(\frac{1}{\delta}\right)
\end{gather*}
is given as
\begin{gather*}
    \frac{t\epsilon^2}{2p_1} \geq \ln\left(\frac{1}{\delta}\right) \\
    \implies t \geq \frac{2p_1}{\epsilon^2}\ln\left(\frac{1}{\delta}\right).
\end{gather*}
To find a sufficient condition for $t$ to satisfy the above, we can find a lower bound on $\epsilon^2$ and use it in the expression above. We have 
\begin{gather*}
    \epsilon^2 =  \left(\left(p_1-\frac{1}{2}\right) -\sqrt{\frac{\log_2\left(\frac{t+1}{\delta}\right)}{2t}} \right)^2\\
    = \left(p_1-\frac{1}{2}\right)^2 + \frac{\log_2\left(\frac{t+1}{\delta}\right)}{2t} - 2\left(p_1-\frac{1}{2}\right)\sqrt{\frac{\log_2\left(\frac{t+1}{\delta}\right)}{2t}}\\
    \implies \epsilon^2 \geq \left(p_1-\frac{1}{2}\right)^2 - 2\left(p_1-\frac{1}{2}\right)\sqrt{\frac{\log_2\left(\frac{t+1}{\delta}\right)}{2t}}.
\end{gather*}
\subsubsection{Constant Factor Tightening}
\label{appA:3}
To obtain our final bound, we 
consider two cases: $p_1 \leq 0.6, p_1 > 0.6$. 
\paragraph{Case 1: \texorpdfstring{$p_1 \leq 0.6$}{}} 
We will find a sufficient condition on $t$ for the fact that 
\begin{gather}
    \frac{t}{C_1}\left(p_1-\frac{1}{2}\right)^2 \geq \log_2\left(\frac{t+1}{\delta}\right)
    \label{appA:eq1}
\end{gather}
where $C_1 = 2.4488$.\\
Earlier in the proof, we have used the fact that $\log_2\left(\frac{\delta}{t + 1}\right) + t \geq 0$. We first prove this fact.
For $p_{1} \in (\frac{1}{2}, 1]$, $\frac{\left(p_1-\frac{1}{2}\right)^2}{C_1} \leq 1$. So, from  \eqref{appA:eq1}, we can easily see that $\log_2\left(\frac{\delta}{t + 1}\right) + t \geq 0$. \\
We find a sufficient $t$ such that 
\begin{gather*}
    \frac{t}{C_1}\left(p_1-\frac{1}{2}\right)^2 \geq \log_2\left(\frac{t+1}{\delta}\right) \\ 
    \implies t \geq \frac{C_1}{(p_1-\frac{1}{2})^2}\log_2\left(\frac{t+1}{\delta}\right).
\end{gather*}
We let $t$ be of the form
\begin{gather*}
    t = \frac{C_1C_2}{(p_1-\frac{1}{2})^2}\log_2\left(\frac{C_1}{(p_1-\frac{1}{2})^2\delta}\right),
\end{gather*}
where $C_2$ is a constant. Note that, for this choice of $t$, (We will show later that $t \geq 4000$ is also true for the first line below to hold),
\begin{gather*}
    t+1 \leq 1.00025t. \\ 
    \frac{C_1}{(p_1-\frac{1}{2})^2}\log_2\left(\frac{t+1}{\delta}\right) \leq \frac{C_1}{(p_1-\frac{1}{2})^2}\log_2\left(\frac{1.00025t}{\delta}\right). \\
    \frac{C_1}{(p_1-\frac{1}{2})^2}\log_2\left(\frac{t+1}{\delta}\right) \leq \frac{C_1}{(p_1-\frac{1}{2})^2}\log_2(1.00025C_2) + 0.3839\frac{C_1}{(p_1-\frac{1}{2})^2}\log_2\left(\frac{C_1}{(\theta-\frac{1}{2})^2\delta}\right) + \\ \frac{C_1}{(p_1-\frac{1}{2})^2}\log_2\left(\frac{C_1}{(p_1-\frac{1}{2})^2\delta}\right),
\end{gather*}
where the last step is true because
\begin{gather*}
    \log_2\left(\log_2\left(\frac{C_1}{(p_1-\frac{1}{2})^2\delta}\right)\right) \leq 0.3839\log_2\left(\frac{C_1}{(p_1-\frac{1}{2})^2\delta}\right)
\end{gather*}
using that, for $x \geq 200$, 
\begin{gather*}
    \log_2\left(\log_2\left(x\right)\right) \leq 0.3839\log_2\left(x\right).
\end{gather*}
Thus, we get 
\begin{gather*}
    \frac{C_1}{(p_1-\frac{1}{2})^2}\log_2\left(\frac{t+1}{\delta}\right) \leq \frac{C_1}{(p_1-\frac{1}{2})^2}\log_2\left(\frac{C_1}{(p_1-\frac{1}{2})^2\delta}\right)\left(1.3839+\log_2\left(1.00025C_2\right)\right) \\ \leq \frac{C_1}{(p_1-\frac{1}{2})^2}\log_2\left(\frac{C_1}{(p_1-\frac{1}{2})^2\delta}\right)C_2
\end{gather*}
where the last step is true when $C_2 = 2.9402$. Thus,
\begin{gather*}
    t = \frac{2C_1C_2p_1}{(p_1-\frac{1}{2})^2}\log_2\left(\frac{C_1}{(p_1-\frac{1}{2})^2\delta}\right)
    \implies t = \frac{C_3p_1}{(p_1-\frac{1}{2})^2}\ln\left(\frac{C_1}{(p_1-\frac{1}{2})^2\delta}\right),
\end{gather*}
where $C_3 = 20.775 \geq 2C_1C_2\log_2(e)$
satisfies the above inequality. \\\\
So now, if the condition on $t$ is satisfied, we have
\begin{gather*}
    \epsilon^2 \geq \left(p_1-\frac{1}{2}\right)^2 - 2\left(p_1-\frac{1}{2}\right)\sqrt{\frac{\log_2\left(\frac{t+1}{\delta}\right)}{2t}} \geq \left(p_1-\frac{1}{2}\right)^2 - 2\left(p_1-\frac{1}{2}\right)\frac{\left(p_1-\frac{1}{2}\right)}{2.21305}\\
    \implies \epsilon^2 \geq C_4\left(p_1-\frac{1}{2}\right)^2,
\end{gather*}
where $C_4 = 0.09627$.\\
Previously, we had remarked that it was sufficient to have 
\begin{gather*}
    t \geq \frac{2p_1}{\epsilon^2}\ln\left(\frac{1}{\delta}\right).
\end{gather*}
We can loosen this by using the lower bound on $\epsilon$ to: it is sufficient to satisfy
\begin{gather*}
    t \geq \frac{C_5p_1}{\left(p_1-\frac{1}{2}\right)^2}\ln\left(\frac{1}{\delta}\right),
\end{gather*}
where $C_5 = 20.775$. \\
To make both the lower bound on $\epsilon$ and the above equation valid, we note that the following choice of $t$ works:
\begin{gather*}
    t = \frac{C_5p_1}{\left(p_1-\frac{1}{2}\right)^2}\max\left(\left(\ln\left(\frac{C_1}{(p_1-\frac{1}{2})^2\delta}\right), \ln\left(\frac{1}{\delta}\right)\right)\right)
\end{gather*}
We note that this leaves us with the following choice of $t$:
\begin{gather*}
    t = \frac{20.775p_1}{\left(p_1-\frac{1}{2}\right)^2}\left(\ln\left(\frac{2.45}{(p_1-\frac{1}{2})^2\delta}\right)\right)
\end{gather*}
Note that $t \geq 4000$, since $p_1 \leq 0.6$.

\paragraph{Case 2: \texorpdfstring{$p_1 > 0.6$}{}}
We will find a sufficient condition on $t$ for the fact that 
\begin{gather*}
    \frac{t}{C_1}\left(p_1-\frac{1}{2}\right)^2 \geq \log_2\left(\frac{t+1}{\delta}\right)
\end{gather*}
where $C_1 = 2.4877$.\\
Using the same argument as in case 1, here also we can prove that $\log_2\left(\frac{\delta}{t + 1}\right) + t \geq 0$.\\
We find a sufficient $t$ such that 
\begin{gather*}
    \frac{t}{C_1}\left(p_1-\frac{1}{2}\right)^2 \geq \log_2\left(\frac{t+1}{\delta}\right) \\ 
    \implies t \geq \frac{C_1}{(p_1-\frac{1}{2})^2}\log_2\left(\frac{t+1}{\delta}\right).
\end{gather*}
We let $t$ be of the form
\begin{gather*}
    t = \frac{C_1C_2}{(p_1-\frac{1}{2})^2}\log_2\left(\frac{C_1}{(p_1-\frac{1}{2})^2\delta}\right),
\end{gather*}
where $C_2$ is a constant. \\
Note that, for this choice of $t$, (We will show later that $t \geq 100$ is also true for the first line to hold)
\begin{gather*}
    t+1 \leq 1.01t. \\ 
    \frac{C_1}{(p_1-\frac{1}{2})^2}\log_2\left(\frac{t+1}{\delta}\right) \leq \frac{C_1}{(p_1-\frac{1}{2})^2}\log_2\left(\frac{1.01t}{\delta}\right). \\
    \frac{C_1}{(p_1-\frac{1}{2})^2}\log_2\left(\frac{t+1}{\delta}\right) \leq \frac{C_1}{(p_1-\frac{1}{2})^2}\log_2(1.01C_2) + 0.5228\frac{C_1}{(p_1-\frac{1}{2})^2}\log_2\left(\frac{C_1}{(\theta-\frac{1}{2})^2\delta}\right) + \\ \frac{C_1}{(p_1-\frac{1}{2})^2}\log_2\left(\frac{C_1}{(p_1-\frac{1}{2})^2\delta}\right),
\end{gather*}
where the last step is true because
\begin{gather*}
    \log_2\left(\log_2\left(\frac{C_1}{(p_1-\frac{1}{2})^2\delta}\right)\right) \leq 0.5228\log_2\left(\frac{C_1}{(p_1-\frac{1}{2})^2\delta}\right)
\end{gather*}
using that, for $x \geq 9.64$, 
\begin{gather*}
    \log_2\left(\log_2\left((x\right)\right) \leq 0.5228\log_2\left(x\right)
\end{gather*}
Thus, we get 
\begin{gather*}
    \frac{C_1}{(p_1-\frac{1}{2})^2}\log_2\left(\frac{t+1}{\delta}\right) \leq \frac{C_1}{(p_1-\frac{1}{2})^2}\log_2\left(\frac{C_1}{(p_1-\frac{1}{2})^2\delta}\right)\left(1.5228+\log_2\left(1.01C_2\right)\right) \\ \leq \frac{C_1}{(p_1-\frac{1}{2})^2}\log_2\left(\frac{C_1}{(p_1-\frac{1}{2})^2\delta}\right)C_2,
\end{gather*}
where the last step is true when $C_2 = 3.228$. Thus, using that $p_1 > 0.6$ by writing that $\frac{p_1}{0.6} \geq 1$, we have that
\begin{gather*}
    t = \frac{1.67C_1C_2p_1}{(p_1-\frac{1}{2})^2}\log_2\left(\frac{C_1}{(p_1-\frac{1}{2})^2\delta}\right)
    \implies t = \frac{C_3p_1}{(p_1-\frac{1}{2})^2}\ln\left(\frac{C_1}{(p_1-\frac{1}{2})^2\delta}\right),
\end{gather*}
where $C_3 = 19.35 \geq 1.67C_1C_2\log_2(e)$
satisfies the above inequality. \\\\
So now, if the condition on $t$ is satisfied, we have
\begin{gather*}
    \epsilon^2 \geq \left(p_1-\frac{1}{2}\right)^2 - 2\left(p_1-\frac{1}{2}\right)\sqrt{\frac{\log_2\left(\frac{t+1}{\delta}\right)}{2t}} \geq \left(p_1-\frac{1}{2}\right)^2 - 2\left(p_1-\frac{1}{2}\right)\frac{\left(p_1-\frac{1}{2}\right)}{2.2305}\\
    \implies \epsilon^2 \geq C_4\left(p_1-\frac{1}{2}\right)^2,
\end{gather*}
where $C_4 = 0.1033$.\\
Previously, we had remarked that it was sufficient to have
\begin{gather*}
    t \geq \frac{2p_1}{\epsilon^2}\ln\left(\frac{1}{\delta}\right).
\end{gather*}
We can loosen this by using the lower bound on $\epsilon$ to: it is sufficient to satisfy
\begin{gather*}
    t \geq \frac{C_5p_1}{\left(p_1-\frac{1}{2}\right)^2}\ln\left(\frac{1}{\delta}\right),
\end{gather*}
where $C_5 = 19.36$. \\
To make both the lower bound on $\epsilon$ and the above equation valid, we note that the following choice of $t$ works:
\begin{gather*}
    t = \frac{C_5p_1}{\left(p_1-\frac{1}{2}\right)^2}\max\left(\left(\ln\left(\frac{C_1}{(p_1-\frac{1}{2})^2\delta}\right), \ln\left(\frac{1}{\delta}\right)\right)\right).
\end{gather*}
We note that this leaves us with the following choice of $t$:

\begin{gather*}
    t = \frac{19.36p_1}{\left(p_1-\frac{1}{2}\right)^2}\left(\ln\left(\frac{2.49}{(p_1-\frac{1}{2})^2\delta}\right)\right)
\end{gather*}
Note that $t \geq 106$.

Combining both the above bounds, we have that
\begin{gather*}
    t = \frac{19.36p_1}{\left(p_1-\frac{1}{2}\right)^2}\left(\ln\left(\frac{2.49}{(p_1-\frac{1}{2})^2\delta}\right)\right)
\end{gather*}
is sufficient.\\\\
Looking over both cases, we note
\begin{gather*}
    t = \frac{20.775p_1}{\left(p_1-\frac{1}{2}\right)^2}\left(\ln\left(\frac{2.49}{(p_1-\frac{1}{2})^2\delta}\right)\right)
\end{gather*}
is sufficient. We denote $C_6 = 20.775, C_7 = 2.49$ to get
\begin{gather*}
    t = \frac{C_6p_1}{\left(p_1-\frac{1}{2}\right)^2}\left(\ln\left(\frac{C_7}{(p_1-\frac{1}{2})^2\delta}\right)\right)
\end{gather*}
as an upper bound on the sample complexity of PPR-Bernoulli with probability atleast $1 - \delta$.
\end{proof}

\subsection{An Upper Bound for General K}
\label{app:ubgenk}

\begin{theorem}[PPR-1v1 upper bound]
\label{thm:ppr-meupperbound}
Fix $\delta \in (0, 1)$, $K \geq 2$, and problem instance $\mathcal{P} = (p, v, K)$. When PPR-1v1 is run on $\mathcal{P}$, with probability $1 - \delta$, the number of samples it observes is at most $$t^{\star} = \frac{194.07p_1}{(p_1-p_2)^2}\ln\left(\sqrt{\frac{79.68(K-1)}{\delta}}\frac{p_1}{(p_1-p_2)}\right).$$
\end{theorem}
\begin{proof}
Fix arbitrary $t > t^{\star}$ and $j \geq 2$. Our strategy is to show that with sufficiently high probability, $v_{1}$ and $v_{j}$ will have separated before $t$ pulls.  First, let $s^t_{ij}$ denote the sum of the number of occurrences of $v_{1}$ and $v_{j}$ in the first $t$ samples: that is, $s^{t}_{ij} = s^{t}_{1} + s^{t}_{j}$. Clearly $s^t_{ij}$ is a binomial random variable with parameters $t$ and $(p_{1} + p_{j})$, We argue that $s^t_{ij}$ cannot fall too far below its mean. Concretely, take $\delta^{\prime} = \frac{\delta}{2 (K - 1)}$ and $l = \sqrt{\frac{2 \ln(\frac{1}{\delta^{\prime}})}{(p_{1} + p_{2})t}}$. A Chernoff bound yields
$$\mathbb{P}\{s^{t}_{ij} \leq (1 - l) (p_{1} + p_{j}) t \} \leq \exp\left(-\frac{l^{2}(p_{1} + p_{j})t}{2}\right) \leq \delta^{\prime}$$
for our choice of $l$. Thus, with probability at least $1 - \delta^{\prime}$, $v_{1}$ and $v_{j}$ together have more than $(1 - l) (p_{1} + p_{j}) t$ samples. Now, the test to separate $v_{1}$ and $v_{j}$ is PPR-Bernoulli on a Bernoulli variable with parameter $q_{1} = \frac{p_{1}}{p_{1} + p_{j}} > \frac{1}{2}$. Although PPR-1v1 runs this test with mistake probability $\frac{\delta}{K - 1}$, imagine running it with mistake probability $\delta^{\prime} = \frac{\delta}{2(K - 1)}$. The latter test would necessarily incur equal or more samples on \textit{every} run. Yet, from Lemma~\ref{lem:ppr-bernoulli-upperbound}, we know that with probability at least $1 - \delta^{\prime}$, the latter will terminate after at most $u = \frac{20.775 q_1}{(q_1-\frac{1}{2})^2}\ln\left(\frac{2.49}{(q_1-\frac{1}{2})^2 \delta^{\prime}}\right)$ samples. It can be verified that for $t > t^{\star}$, $u < (1 - l)(p_{1} + p_{j})t$ (calculation shown in Appendix~\ref{app:calculationinproofoftheorempprmeupperbound}). In other words, the probability that $v_{1}$ and $v_{j}$ have \textit{not} separated before $t$ pulls is at most $2\delta^{\prime}$.

Since the argument above holds for arbitrary $j \geq 2$, a union bound establishes that with probability at least $1 - \delta$, $v_{1}$ must have separated from all other values---implying termination---before $t$ pulls.
\end{proof}

\subsubsection{Calculation in Proof of Theorem~\ref{thm:ppr-meupperbound}} \label{app:calculationinproofoftheorempprmeupperbound}

We have to show $u < (1 - l)(p_{1} + p_{j})t$ for $t > t^{\star}$, where
\begin{align*}
u &= \frac{20.775 q_1}{(q_1-\frac{1}{2})^2}\ln\left(\frac{2.49}{(q_1-\frac{1}{2})^2 \delta^{\prime}}\right), \\
q_{1} &= \frac{p_{1}}{p_{1} + p_{j}},\\
l &= \sqrt{\frac{2 \ln(1/\delta^{\prime})}{(p_{1} + p_{j})t}}, \text{ and}\\
t^{\star} &= \frac{194.07p_1}{(p_1-p_2)^2}\ln\left(\sqrt{\frac{79.68(k-1)}{\delta}}\frac{p_1}{(p_1-p_2)}\right).
\end{align*}
First, observe that
\begin{equation*}
u = \frac{83.1p_1(p_1+p_j)}{(p_1-p_j)^2}\ln\left(\frac{9.96(p_1+p_j)^2}{(p_1-p_j)^2 \delta^{\prime}}\right) \leq \frac{83.1p_1(p_1+p_j)}{(p_1-p_2)^2}\ln\left(\frac{39.84p_1^2}{(p_1-p_2)^2 \delta^{\prime}}\right).
\end{equation*}
Thus, we have
\begin{gather*}
    \frac{u}{(p_1+p_j)t^*} < 0.85640\frac{\ln\left(\frac{39.84p_1^2}{(p_1-p_2)^2 \delta^{\prime}}\right)}{\ln\left(\frac{79.68(k-1)}{\delta}\frac{p_1^2}{(p_1-p_2)^2}\right)} = 0.85640.
\end{gather*}
It suffices to show that $l \leq 0.1436$, which is established by the steps below.
\begin{align*}
l^2 &\leq \frac{2\ln\left(1/\delta^{\prime}\right)(p_1-p_2)^2}{(p_1+p_j)97.035p_1\ln\left(\frac{79.68(k-1)p_1^2}{\delta (p_1-p_2)^2}\right)} \leq \frac{2(p_1-p_2)^2}{(p_1+p_j)97.035p_1} \leq \frac{2}{97.035} \\
&\implies l \leq \frac{1}{\sqrt{97.035/2}} < 0.1436.
\end{align*}

\newpage

%% file: appendices/dcb.tex
\newpage
\section{DCB ALGORITHM}
\label{app:dcbalgorithm}

We describe the DCB (``Difference in Confidence Bounds'') algorithm for sampling constituencies in the indirect election winner-forecasting problem from Section~\ref{sec:elections}.

Suppose we have $C$ constituencies. For ease of explanation, we assume that all the constituencies have the same set of $K \geq 2$ parties (in reality we maintain a separate list for each constituency). Although constituencies have finite populations, these are usually large enough to ignore the benefit of without-replacement samples; we simply view each response as a sample from a discrete distribution.

After $t \geq 1$ samples have been obtained from the population, let $\text{LCB}^{t}(c, i)$ and $\text{UCB}^{t}(c, i)$ denote (1vr) lower and upper confidence bounds, respectively, on the the (true) fraction of votes to be cast for party $i \in \{1, 2, \dots ,K\}$ in constituency $c \in \{1, 2, \dots, C\}$. If applying a 1v1 procedure for mode estimation, we have separate confidence bounds $\text{LCB}^{t}(c, i, j)$ and 
$\text{UCB}^{t}(c, i, j)$ for each pair of parties $i, j \in \{1, 2, \dots, K\}$, $i \neq j$. The permitted mistake probability $\delta$ is divided equally among the constituencies.

The key idea in the DCB algorithm is to keep lower and upper confidence bounds on the \textit{wins of each party} (across constituencies), and to use this information to guide sampling. For party $i\in \{1, 2, \dots, K\}$, the current number of wins ($\text{wins}^{t}_{i}$) and losses ($\text{losses}^{t}_{i}$) can be obtained by verifying whether its confidence bounds within each constituency have separated accordingly from other parties. We also use $\text{leads}^{t}_{i}$ to denote the number of constituencies in which $i$ has polled the most votes yet, but has not yet won. Thus, we have $\text{LCB}^{t}_{i} = \text{wins}^{t}_{i} \text{, and } \text{UCB}^{t}_{i} = C - 
\text{losses}^{t}_{i}.$

Observe that the overall winner can be declared as soon as one party's LCB exceeds the UCB of all the other parties. On the other hand, when the winner is yet to be identified, one would ideally like to focus on ``potential'' winners, rather than query a constituency whose result does not seem relevant to
the big picture. Taking cue from the LUCB algorithm for bandits~\cite{ICML12-shivaram}, the first step under DCB is to identify two contenders for the top position: party $a^{t} = \argmax_{i \in \{1, 2, \dots, K\}} (\text{wins}^{t}_{i} + \text{leads}^{t}_{i})$, and party $b^{t} = \argmax_{i \in \{1, 2, \dots, K\} \setminus \{a^{t}\}} \text{UCB}^{t}_{i}$. Optimistic that sampling can reveal a win for $a^{t}$ and a loss for $b^{t}$---which would take us closer to termination---DCB picks one ``promising'' constituency each for $a^{t}$ and $b^{t}$. These constituencies, denoted $c^{t}_{1}$ and $c^{t}_{2}$, are defined below for use with both  1v1 and 1vr confidence bounds. The idea is the same: $c^{t}_{1}$ is the constituency in which $a^{t}$ appears poised to win by a large margin, and $c^{t}_{2}$ is the constituency in which $b^{t}$ appears poised to lose by a large margin.
\begin{align*}
c^{t}_{1} &=
\begin{cases}
\argmax_{c} \min_{j \in \{1, 2, \dots, K\}} (\text{UCB}(c, a^{t}, j) - \text{LCB}(c, a^{t}, j)) & \text{(1v1),} \\
\argmax_{c} \min_{j \in \{1, 2, \dots, K\}} (\text{UCB}(c, a^{t}) - \text{LCB}(c, j)) & \text{(1vr),}
\end{cases}\\
c^{t}_{2} &=
\begin{cases}
\argmax_{c} \max_{j \in \{1, 2, \dots, K\}} (\text{UCB}(c, j, b^{t}) - \text{LCB}(c, j, b^{t})) & \text{(1v1),} \\
\argmax_{c} \max_{j \in \{1, 2, \dots, K\}} (\text{UCB}(c, j) - \text{LCB}(c, b^{t})) & \text{(1vr).}
\end{cases}
\end{align*}
The outer ``$\argmax$'' in all the definitions above is over all constituencies where the concerned party ($a^t$ or $b^t$) is still in contention: that is, it has not yet won or lost the constituency. The DCB algorithm queries $c^{t}_{1}$ and $c^{t}_{2}$ at each time step $t$, and terminates once an overall winner has been identified.

\phantom{aaaa}

\phantom{aaaa}

\phantom{aaaa}

\phantom{aaaa}

\phantom{aaaa}

\phantom{aaaa}

\phantom{aaaa}

\phantom{aaaa}

\phantom{aaaa}

%% file: appendices/bihar-table.tex
\newpage
\section{Winner Forecasting in Bihar Elections}
\label{app:bihar-table}

Below we provide results from the 2015 Bihar state elections, a closer contest than the 2014 Indian national elections whose results were shown in Table~\ref{tab:elections-india} in Section~\ref{sec:elections}. Of a total 242 seats in this election, 80 went to the winner, 71 to the second largest party, and 53 to the third largest.

\begin{table}[H]
\footnotesize
\centering
\scshape
  \caption{Sample complexity of various stopping rules when coupled with (1) round-robin (RR) polling of constituencies and (2) DCB. All experiments are run with mistake probability $\delta=0.01$. Values shown are averages from 10 runs, and show one standard error.  ``Seats resolved'' indicates the number of constituencies in which a winner was identified before the overall procedure terminated.}
  \label{tab:elections-bihar}  \begin{tabular}{lccccc}
    \toprule
    \multirow{3}{*}{Algorithm} &
      \multicolumn{2}{c}{Bihar-2015 (242 seats)} \\
      &  \multirow{2}{*}{Samples} & {Seats}\\
      &  & {Resolved}\\
      \midrule
    RR-$\mathcal{A}_1$-1v1 & 4201429 $\pm$ 43932  & 221 $\pm$ 1 \\
    RR-$\mathcal{A}_1$-1vr & 4841406 $\pm$ 58188 & 222 $\pm$ 1 \\
    RR-KLSN-1v1 & 2198678 $\pm$ 44514 & 221 $\pm$ 1 \\
    RR-KLSN-1vr & 2668729 $\pm$ 47535 & 222 $\pm$ 2 \\
    RR-PPR-1v1 &\textbf{ 1813213 $\pm$ 42081 }& 221 $\pm$ 1 \\
    RR-PPR-1vr & 2054171 $\pm$ 40394 & 221 $\pm$ 1 \\\midrule
    DCB-$\mathcal{A}_1$-1v1 & 2301936 $\pm$ 36399 & 135 $\pm$ 1 \\
    DCB-$\mathcal{A}_1$-1vr & 2481552 $\pm$ 21483 & 134 $\pm$ 1 \\
    DCB-KLSN-1v1 & 1127963 $\pm$ 22191 & 139 $\pm$ 2 \\
    DCB-KLSN-1vr & 1312027 $\pm$ 22892 & 139 $\pm$ 1 \\
    DCB-PPR-1v1 & \textbf{883389 $\pm$ 15581} & 142 $\pm$ 2 \\
    DCB-PPR-1vr & 993495 $\pm$ 18859 & 139 $\pm$ 1 \\
    \bottomrule
  \end{tabular}
\normalsize
\end{table}

%% file: main.bbl
\begin{thebibliography}{23}
\providecommand{\natexlab}[1]{#1}
\providecommand{\url}[1]{\texttt{#1}}
\expandafter\ifx\csname urlstyle\endcsname\relax
  \providecommand{\doi}[1]{doi: #1}\else
  \providecommand{\doi}{doi: \begingroup \urlstyle{rm}\Url}\fi

\bibitem[Buterin(2014)]{buterin2014}
Vitalik Buterin.
\newblock A next-generation smart contract and decentralized application
  platform, 2014.
\newblock URL
  \url{https://cryptorating.eu/whitepapers/Ethereum/Ethereum_white_paper.pdf}.

\bibitem[Chen et~al.(2008)Chen, Park, and Bian]{chen2008}
R.~Chen, J.-M. Park, and K.~Bian.
\newblock Robust distributed spectrum sensing in cognitive radio networks.
\newblock In \emph{IEEE INFOCOM 2008 - The 27th Conference on Computer
  Communications}, pages 1876--1884. IEEE Press, 2008.
\newblock \doi{10.1109/INFOCOM.2008.251}.

\bibitem[Das et~al.(2019)Das, Ribeiro, and Anand]{sourav2019}
Sourav Das, Vinay~Joseph Ribeiro, and Abhijeet Anand.
\newblock {YODA:} enabling computationally intensive contracts on blockchains
  with {B}yzantine and {S}elfish nodes.
\newblock In \emph{26th Annual Network and Distributed System Security
  Symposium, {NDSS}}. The Internet Society, 2019.

\bibitem[Galvin(2014)]{galvin2014}
David Galvin.
\newblock Three tutorial lectures on entropy and counting.
\newblock \emph{arXiv preprint arXiv:1406.7872}, 2014.

\bibitem[Garivier(2013)]{garivier2013}
Aurélien Garivier.
\newblock Informational confidence bounds for self-normalized averages and
  applications.
\newblock In \emph{2013 IEEE Information Theory Workshop (ITW)}, pages 1--5.
  IEEE press, 2013.
\newblock \doi{10.1109/ITW.2013.6691311}.

\bibitem[Garivier and Kaufmann(2016)]{pmlr-v49-garivier16a}
Aurélien Garivier and Emilie Kaufmann.
\newblock Optimal best arm identification with fixed confidence.
\newblock In \emph{29th Annual Conference on Learning Theory}, volume~49 of
  \emph{Proceedings of Machine Learning Research}, pages 998--1027. PMLR, 2016.

\bibitem[Gross and Humenik(1991)]{kenny1991}
Kenny~C. Gross and Keith~E. Humenik.
\newblock Sequential probability ratio test for nuclear plant component
  surveillance.
\newblock \emph{Nuclear Technology}, 93\penalty0 (2):\penalty0 131--137, 1991.
\newblock \doi{10.13182/NT91-A34499}.

\bibitem[Haddenhorst et~al.(2021)Haddenhorst, Bengs, and
  H{\"u}llermeier]{haddenhorst}
Bj{\"o}rn Haddenhorst, Viktor Bengs, and Eyke H{\"u}llermeier.
\newblock Identification of the generalized {C}ondorcet winner in multi-dueling
  bandits.
\newblock In A.~Beygelzimer, Y.~Dauphin, P.~Liang, and J.~Wortman Vaughan,
  editors, \emph{Advances in Neural Information Processing Systems}, 2021.
\newblock URL \url{https://openreview.net/forum?id=omDF-uQ_OZ}.

\bibitem[Howard et~al.(2020)Howard, Ramdas, McAuliffe, and Sekhon]{Howard20}
Steven~R. Howard, Aaditya Ramdas, Jon McAuliffe, and Jasjeet Sekhon.
\newblock Time-uniform {C}hernoff bounds via nonnegative supermartingales.
\newblock \emph{Probability Surveys}, 17:\penalty0 257--317, 2020.

\bibitem[Jamieson et~al.(2014)Jamieson, Malloy, Nowak, and Bubeck]{Jamieson14}
Kevin Jamieson, Matthew Malloy, Robert Nowak, and Sébastien Bubeck.
\newblock lil' ucb : An optimal exploration algorithm for multi-armed bandits.
\newblock In \emph{Proceedings of The 27th Conference on Learning Theory},
  volume~35 of \emph{Proceedings of Machine Learning Research}, pages 423--439.
  PMLR, 2014.

\bibitem[Kalyanakrishnan et~al.(2012)Kalyanakrishnan, Tewari, Auer, and
  Stone]{ICML12-shivaram}
Shivaram Kalyanakrishnan, Ambuj Tewari, Peter Auer, and Peter Stone.
\newblock {PAC} subset selection in stochastic multi-armed bandits.
\newblock In \emph{Proceedings of the 29th International Conference on
  International Conference on Machine Learning}, page 227–234. Omnipress,
  2012.

\bibitem[Karandikar(2018)]{karandikar2018}
Rajeeva Karandikar.
\newblock Power and limitations of opinion polls in the context of {I}ndian
  parliamentary democracy.
\newblock In \emph{Special Proceeding of 20th Annual Conference of SSCA}, pages
  09 -- 16. Society of Statistics, Computer and Applications, 2018.

\bibitem[Karandikar et~al.(2002)Karandikar, Payne, and Yadav]{rajeeva2002}
Rajeeva~L. Karandikar, Clive Payne, and Yogendra Yadav.
\newblock Predicting the 1998 {I}ndian parliamentary election.
\newblock \emph{Electoral Studies}, 21\penalty0 (1):\penalty0 69--89, 2002.
\newblock \doi{https://doi.org/10.1016/S0261-3794(00)00042-1}.

\bibitem[Kaufmann and Kalyanakrishnan(2013)]{pmlr-v30-Kaufmann13}
Emilie Kaufmann and Shivaram Kalyanakrishnan.
\newblock Information complexity in bandit subset selection.
\newblock In \emph{Proceedings of the 26th Annual Conference on Learning
  Theory}, volume~30 of \emph{Proceedings of Machine Learning Research}, pages
  228--251. PMLR, 2013.

\bibitem[Manku and Motwani(2002)]{Motwani02}
Gurmeet~Singh Manku and Rajeev Motwani.
\newblock Approximate frequency counts over data streams.
\newblock In \emph{Proceedings of the 28th International Conference on Very
  Large Data Bases}, page 346–357. VLDB Endowment, 2002.

\bibitem[Maurer and Pontil(2009)]{maurer2009}
Andreas Maurer and Massimiliano Pontil.
\newblock Empirical {B}ernstein bounds and sample variance penalization.
\newblock In \emph{Proceedings of the 22nd Conference on Learning Theory (COLT
  2009)}, 2009.
\newblock URL \url{http://www.cs.mcgill.ca/~colt2009/papers/012.pdf\#page=1}.

\bibitem[Mulzer(2018)]{mulzer2019proofs}
Wolfgang Mulzer.
\newblock Five proofs of {C}hernoff's bound with applications.
\newblock \emph{CoRR}, abs/1801.03365, 2018.

\bibitem[Nakamoto(2009)]{NakamotoBitcoin}
Satoshi Nakamoto.
\newblock Bitcoin: A peer-to-peer electronic cash system, 2009.
\newblock URL \url{http://www.bitcoin.org/bitcoin.pdf}.

\bibitem[Parzen(1962)]{Parzen62}
Emanuel Parzen.
\newblock On estimation of a probability density function and mode.
\newblock \emph{The Annals of Mathematical Statistics}, 33\penalty0
  (3):\penalty0 1065--1076, 1962.

\bibitem[Payne(2003)]{payne2003}
Clive Payne.
\newblock Election forecasting in the {UK}: The {BBC}'s experience.
\newblock \emph{Euramerica}, 33\penalty0 (1):\penalty0 193--234, 2003.

\bibitem[Shah et~al.(2020)Shah, Choudhury, Karamchandani, and
  Gopalan]{Shah_Choudhury_Karamchandani_Gopalan_2020}
Dhruti Shah, Tuhinangshu Choudhury, Nikhil Karamchandani, and Aditya Gopalan.
\newblock Sequential mode estimation with oracle queries.
\newblock In \emph{Proceedings of the Thirty-Fourth AAAI Conference on
  Artificial Intelligence}, volume~34, pages 5644--5651. AAAI Press, 2020.

\bibitem[Wald(1945)]{awald1945}
A.~Wald.
\newblock {Sequential Tests of Statistical Hypotheses}.
\newblock \emph{The Annals of Mathematical Statistics}, 16\penalty0
  (2):\penalty0 117--186, 1945.

\bibitem[Waudby-Smith and Ramdas(2020)]{waudby-ramdas-ppr}
Ian Waudby-Smith and Aaditya Ramdas.
\newblock Confidence sequences for sampling without replacement.
\newblock In \emph{Advances in Neural Information Processing Systems},
  volume~33, pages 20204--20214. Curran Associates, Inc., 2020.

\end{thebibliography}
